\documentclass[acmlarge]{acmart}
\usepackage{amsmath,amsfonts}
\usepackage{graphicx}
\usepackage{subcaption}
\usepackage{multirow}
\usepackage{makecell}
\usepackage{array}
\usepackage{tabularx}
\usepackage[table]{xcolor}
\definecolor{lightgreen}{HTML}{B6DFB4}
\usepackage{textcomp}
\usepackage{pifont}
\usepackage[linesnumbered,ruled,vlined]{algorithm2e}

\AtBeginDocument{%
  }

\setcopyright{acmlicensed}
\copyrightyear{2026}
\acmYear{2026}
\acmDOI{XXXXXXX.XXXXXXX}

\acmJournal{TODAES}
\acmVolume{37}
\acmNumber{4}
\acmArticle{111}
\acmMonth{8}

\begin{document}



\title{LightMat-HP: A Photonic–Electronic System for Accelerating General Matrix Multiplication With Configurable Precision}

\author{Hailong Gong}
\affiliation{%
  \institution{School of Computing, Australian National University}
  \city{Canberra}
  \state{ACT}
  \country{Australia}}
\email{hailong.gong@anu.edu.au}

\author{Haibo Zhang}
\affiliation{%
  \institution{School of Computer Science and Engineering, University of New South Wales}
  \city{Sydney}
  \state{NSW}
  \country{Australia}}
\email{haibo.zhang@unsw.edu.au}

\author{Amanda S. Barnard}
\affiliation{%
  \institution{School of Computing, Australian National University}
  \city{Canberra}
  \state{ACT}
  \country{Australia}}
\email{amanda.s.barnard@anu.edu.au}

\author{Mahbub Hassan}
\affiliation{%
  \institution{School of Computer Science and Engineering, University of New South Wales}
  \city{Sydney}
  \state{NSW}
  \country{Australia}}
\email{mahbub.hassan@unsw.edu.au}

\author{Matt Woolley}
\affiliation{%
  \institution{School of Engineering and Information Technology, University of New South Wales, Australian Defence Force Academy}
  \city{Canberra}
  \state{ACT}
  \country{Australia}}
\email{m.woolley@unsw.edu.au}

\author{Rajkumar Buyya}
\affiliation{%
  \institution{School of Computing and Information Systems, The University of Melbourne}
  \city{Melbourne}
  \state{VIC}
  \country{Australia}}
\email{rbuyya@unimelb.edu.au}

\renewcommand{\shortauthors}{GONG et al.}

\begin{abstract}
Matrix multiplication is a fundamental kernel in large-scale artificial intelligence and scientific computing, but its performance on conventional electronic accelerators is increasingly constrained by memory bandwidth and energy efficiency. Photonic computing offers a promising alternative due to its ultra-high bandwidth, massive parallelism, and low power dissipation. However, most existing photonic systems are limited to low-precision computation because of analog optical modulation constraints and noise accumulation, which restricts their applicability in precision-critical workloads. To address this limitation, we propose LightMat-HP, a hybrid photonic–electronic computing system that enables end-to-end acceleration of general matrix multiplication with configurable computational precision. LightMat-HP adopts block floating-point (BFP) arithmetic to reduce computational complexity while enabling flexible precision–performance tradeoffs. To overcome the precision limitations of photonic devices, we propose a slicing-based photonic multiplication scheme that exploits the high accuracy of low bit-width photonic multiplication in combination with digital accumulation to achieve high-precision mantissa multiplication. A tile-based matrix multiplication dataflow is further designed to support matrices of arbitrary sizes. We experimentally validate LightMat-HP on a photonic computing prototype and evaluate its performance through large-scale simulations. The results demonstrate that LightMat-HP outperforms FPGA, GPU, and a state-of-the-art photonic accelerator across throughput, latency, and energy efficiency, particularly for small- and medium-sized matrix multiplications, owing to its highly parallel photonic architecture, efficient data movement, and slice-based BFP arithmetic.
These results demonstrate the feasibility and efficiency of photonic–electronic computation and position LightMat-HP as a viable architecture for accelerating matrix operations in future Artificial Intelligence (AI) and scientific computing systems.

\end{abstract}


\ccsdesc[500]{Hardware~Emerging optical and photonic technologies}
\ccsdesc[300]{Computer systems organization~Optical computing}
\ccsdesc[300]{Computing methodologies~Linear algebra algorithms}

\keywords{Photonic Computing, Matrix Multiplication Acceleration, Energy-efficient, High-Performance Systems}

\keywords{Photonic Computing, Matrix Multiplication Acceleration, Energy-efficient, High-Performance Systems}

\received{xx January 2026}
\received[revised]{xx xx 2026}
\received[accepted]{xx xx 2026}
\maketitle

\section{Introduction}

With the rapid development of artificial intelligence (AI), deep neural networks (DNNs) have grown dramatically in scale, leading to a substantial increase in computational throughput and energy demands. This stems from the fact that most computationally intensive operations in DNNs, including fully connected layers, convolution, and attention, can be expressed as matrix multiplication or equivalent linear algebra operations. High model accuracy typically requires large model sizes and high numerical precision, which significantly increases the scale of these matrix operations. As model sizes continue to grow, the computational cost of matrix multiplication increases rapidly, leading to escalating compute and energy demands that increasingly challenge conventional electronic hardware.
This trend is prominent in scenarios such as deep learning training and scientific computing, where workloads often rely on high-precision numerical formats to ensure numerical stability and controllable cumulative errors. These problems reflect multiple structural constraints of the post-Moore era: on the one hand, the slowdown of Moore's Law and the end of Dennard scaling have increasingly limited further miniaturization of transistors in terms of area and power consumption; on the other hand, power density continues to rise, constantly compressing the system's thermal design space. In addition, the inherent \textit{memory wall} problem of the von Neumann architecture further constrains the performance of electronic computing systems. The power consumption of modern supercomputing systems has reached tens of megawatts~\cite{sirbu2016power}, indicating that electronic computing systems are no longer sufficient to achieve a balance between performance, energy efficiency, and sustainability while simultaneously meeting the demands for high-throughput and high-precision computing. Together, these factors reveal a fundamental bottleneck of electronic computing and motivate researchers to explore new computing paradigms capable of supporting energy-efficient, high-precision matrix operations.

\textbf{Photonic computing} has emerged as a promising alternative. By leveraging the high bandwidth, low latency, and inherent parallelism of light, photonic computing promises to overcome the limitations of conventional electronic computing in terms of performance and energy efficiency~\cite{bente2025potential}. More importantly, photonic computing offers a physical-level implementation of large-scale parallel matrix operations, which is totally different from electronic computing. Based on optical modulation, interference, and detection, photonic systems can perform multiplication and addition operations directly in the optical domain, providing a novel computational path for matrix-intensive workloads. Compared with electrical interconnects, optical interconnects generate almost no resistance or capacitance losses during data transmission~\cite{maniyara2016antireflection}, thus offering a significant advantage in energy efficiency. Furthermore, optical channels naturally support multi-wavelength parallelism via wavelength-division multiplexing (WDM)~\cite{yu2025parallel}, enabling multiple operations to be executed simultaneously in a single computation cycle. With the continuous advancement of modern electro-optic materials and device technology, the operating frequencies of photonic devices have reached hundreds of GHz or even THz, far exceeding the GHz frequencies of electronic processors~\cite{lampert2025photonics}.

Although photonic computing offers great potential, several \textbf{challenges} hinder its practical deployment to achieve high-precision computation operations, such as high-precision general matrix multiplication (HP-GEMM). First, photonic computation relies on high-speed analog optical signals and therefore must be tightly coupled with electronic control, buffering, and digitization due to the absence of mature optical memory~\cite{ning2024photonic}. Hence, designing an efficient hybrid photonic–electronic architecture remains a fundamental challenge. Second, most existing photonic accelerators support only low to moderate numerical precision, typically in the range of 4 to 8 bits~\cite{Liu2019HolyLightAN}, which is insufficient for high-precision workloads such as deep learning training and scientific computing. The precision of photonic computing systems is constrained by analog noise, the limited linearity of photonic devices, and the resolution of Digital-to-Analog Converters (DACs) and Analog-to-Digital Converters (ADCs). Unlike electronic computing systems that support a wide range of data types and numerical precisions, including integer and floating-point formats, mixed-precision execution, and custom numeric representations, there is currently a lack of photonic–electronic computing architectures capable of supporting configurable computational precision.


These limitations reveal a clear research gap: \textbf{the absence of hybrid photonic–electronic computing architectures that support end-to-end general matrix multiplication with configurable computational precision}. To address this gap, we design and implement \textbf{LightMat-HP}, an efficient hybrid photonic–electronic system that leverages Block Floating-Point (BFP) arithmetic, operand slicing, and matrix tiling to achieve precision-configurable photonic matrix multiplication. The key contributions of this paper are summarized as follows:


\begin{itemize}
\item We conduct an experimental study on a photonic computing platform to examine the computational accuracy achievable with varying operand bit-widths. Experimental results show that photonic multiplication achieves high computational accuracy when the operand bit-width does not exceed 5 bits, and photonic dot products can achieve relative errors below 1\% by leveraging BFP, slicing-based low-bit photonic multiplication, and  digital accumulation of low-bit photonic products.

\item We design \textbf{LightMat-HP}, a hybrid photonic–electronic computing system for end-to-end acceleration of matrix multiplication. By adopting BFP arithmetic, our system reduces computational complexity and leverages photonic computing to accelerate floating-point mantissa multiplication. We further propose a slicing-based, precision-configurable photonic multiplication scheme that exploits the high accuracy of low bit-width photonic computation to achieve high-precision multiplication for larger bit widths. A tile-based matrix multiplication dataflow is designed to support matrices of arbitrary size.

\item We implement and validate LightMat-HP on our photonic computing prototype and conduct both experiments and large-scale system simulations to evaluate the performance of LightMat-HP. Results demonstrate that LightMat-HP achieves lower latency and improved energy efficiency compared to CPU, GPU, FPGA, and a photonic baseline,  especially for matrices at small and medium sizes. 
\end{itemize}

The remainder of the paper is organized as follows: Section~\ref{sec:background} provides the background of the BFP format and the fundamentals of photonic computing. In Section~\ref{sec:exp_mov}, we experimentally characterize the precision of photonic multiplication and dot-product to motivate the design of LightMat-HP. We then present the complete LightMat-HP system design in Section~\ref{sec:LightMat-HP}. Section~\ref{sec:precision_analysis} models the numerical precision and performs end-to-end error analysis for matrix multiplication in LightMat-HP. Section~\ref{sec:implementation} describes the  methodology for our evaluation. Section~\ref{sec:evaluation} presents the evaluation LightMat-HP on general matrix multiplication. Section~\ref{sec:relatedwork} discusses the related work. Section~\ref{sec:conclusion} concludes this work and discusses future research.

\section{Background}\label{sec:background}

\subsection{Block-Floating-Point Format}\label{sec:bfp}

\begin{figure}[htbp]
  \centering
  \includegraphics[width=0.8\textwidth]{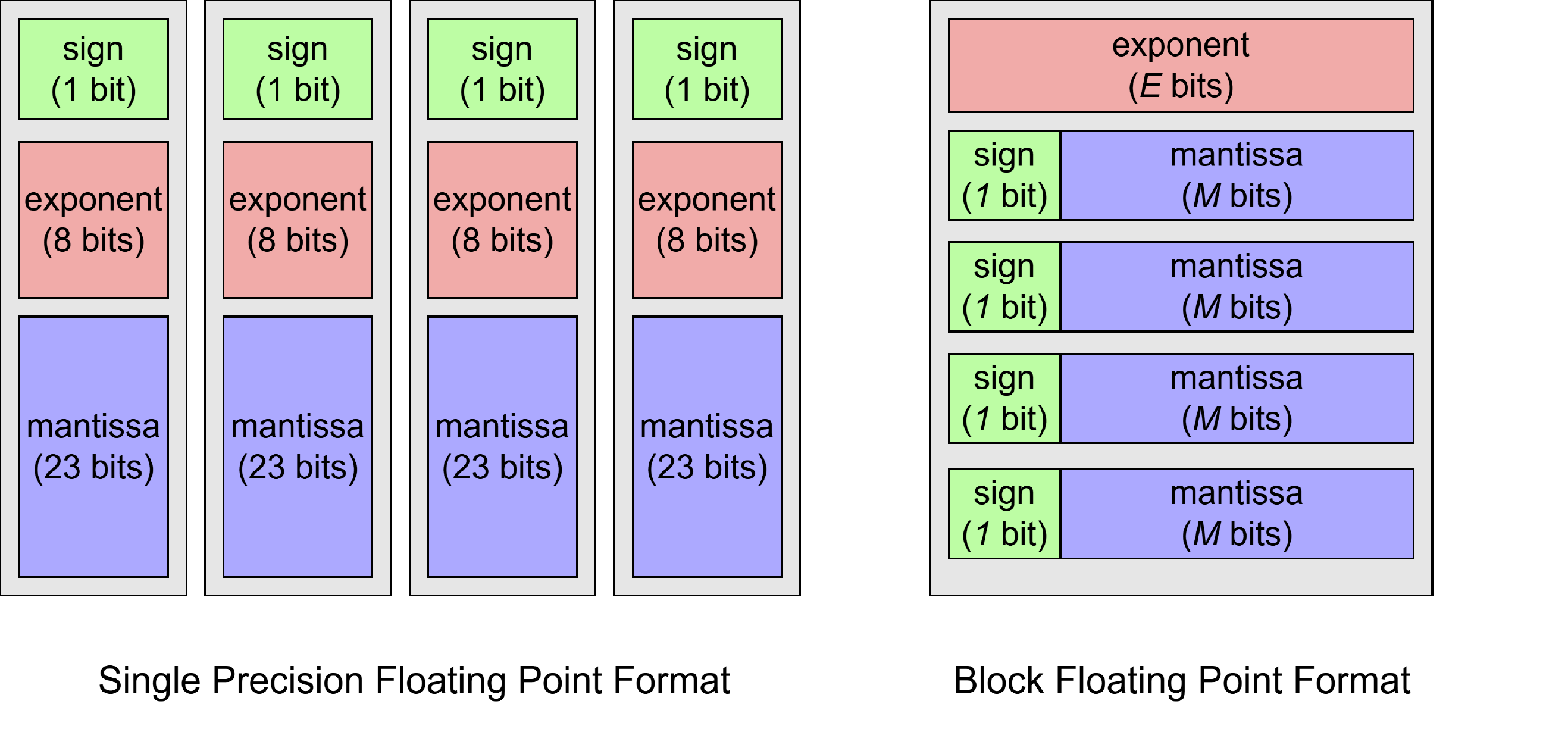}
  \caption{Diagram of Floating Point 32 (FP32) and  Block Floating Point (BFP) Representation.}\label{fig:bfp}
  \Description{}
\end{figure}

The Block-Floating-Point format~\cite{oppenheim2003realization} is a numerical representation that bridges the gap between traditional floating-point (FP) and fixed-point formats. As illustrated in Fig. \ref{fig:bfp}, BFP assigns a single shared exponent to a block of values while preserving individual mantissas, thereby maintaining a wide dynamic range with reduced exponent storage and simplified arithmetic compared to standard floating-point representations. 

Given a block of values $\mathbf{X} = (x_1, x_2, \ldots, x_N)$ in FP format, the conversion to BFP proceeds as follows. First, each value is expressed in normalized floating-point form
\begin{equation}
\mathbf{X} = (x_1, x_2, \ldots, x_N) \\
= (m_1 \times 2^{e_1}, m_2 \times 2^{e_2}, \ldots, m_N \times 2^{e_N}),
\end{equation} where each $x_i$ is represented by its own mantissa $m_i$ and exponent $e_i$. The shared block exponent is then selected as 
\begin{equation}
 e_{s} = \lfloor \log_2(\max_{i=1}^{N}|x_i|)\rfloor-(b-1)
\end{equation}
where $b$ is the mantissa bit-width including the sign bit. This ensures that all values fit within the mantissa range and precision is maximized. Each value $x_i$ is subsequently rescaled to the shared exponent by right-shifting its mantissa by $e_{s} - e_i$ bits $ (
\tilde{m_i} = m_i \times 2^{e_{s} - e_i}$), and then quantized to $b$ bits using rounding or truncation: $\hat{m_i} = Q(\tilde{m})$. Finally, the block $\mathbf{X}$ in BFP format is represented as $\mathbf{X}=(e_s, \{\hat{m_1}, \hat{m_2}, ..., \hat{m_N}\})$.

BFP format offers several advantages for accelerating matrix multiplication in photonic computing: (1) \textit{it significantly reduces exponent-related hardware overhead while maintaining a large dynamic range}. By sharing a single exponent across each block, it avoids exponent arithmetic in the optical domain and enables efficient photonic multiplication using fixed bit-width mantissas. (2) \textit{The mantissa bit width is configurable, providing a flexible tradeoff between precision and performance}. This makes BFP more robust than pure integer quantization for general GEMM workloads. (3) \textit{It naturally matches the tiled execution model of GEMM, allowing exponent reuse within tiles and simplifying accumulation.} Together, these properties reduce the arithmetic complexity, memory bandwidth requirement, and data-movement energy compared to conventional floating-point formats.


\subsection{Fundamentals of Photonic Computing}\label{subsec:principle_photonic}
\begin{figure}[htbp]
  \centering
  \includegraphics[width=0.6\textwidth]{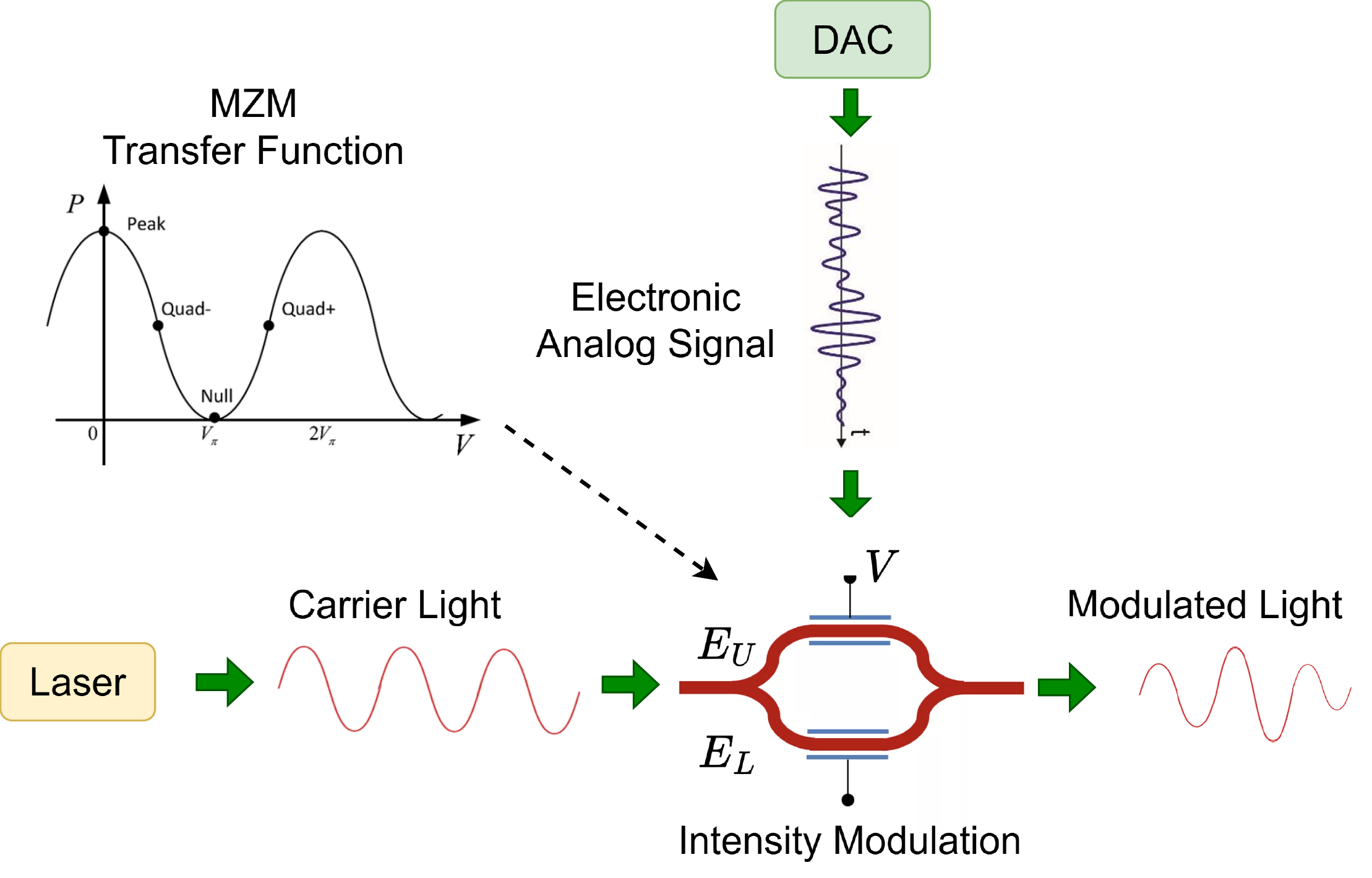}
  \caption{System-level representation of optical intensity modulation using MZM.}\label{fig:mzm}
  \Description{}
\end{figure}

The key principle of photonic computing is to encode digital data onto light using amplitude, phase, or both, and perform computation directly in the optical domain. Mach–Zehnder Modulator (MZM)~\cite{saleh2019fundamentals} is a core photonic device for encoding digital signals onto optical carriers. Fig.~\ref{fig:mzm} illustrates continuous intensity modulation using an MZM. The carrier light is split into two waveguide arms, and the digital data is converted to a Radio Frequency (RF) analog signal, which drives the phase shifters in the two arms and changes their relative phase via the electro-optic effect. When the two paths recombine, their constructive or destructive interference produces the modulated output intensity. The recombined optical field is $E_\text{out} = \frac{1}{\sqrt{2}}(E_\text{U} + E_\text{L})$. The output intensity therefore follows $I_\text{out} \propto \cos^2\!\left(\frac{\Delta\theta}{2}\right)$, where $\Delta\theta = \theta_U - \theta_L$ is the phase difference induced by the differential drive voltage applied to the two arms. According to the MZM intensity transfer function, different input voltages (representing different digital data) induce different phase shifts $\Delta\theta$, thereby modulating the carrier light to produce different output intensities. 

\begin{figure}[htbp]
  \includegraphics[width=0.6\textwidth]{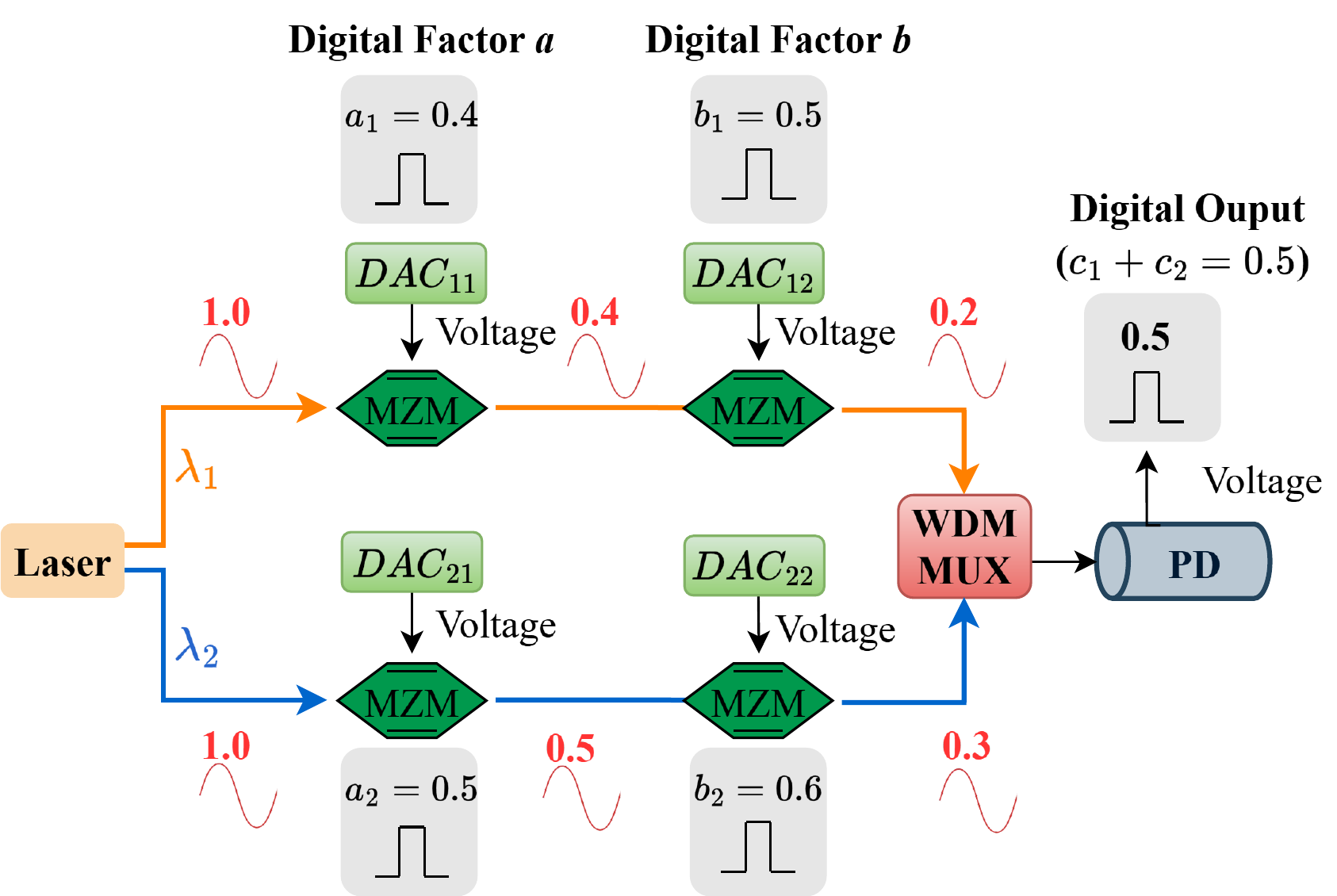}
  \caption{Illustrations of fundamental of photonic computing: photonic multiplication using cascaded MZMs, and photonic addition using MZMs and a WDM combiner.}\label{fig:photonic_operations}
  \Description{}
\end{figure}%

\textbf{Photonic Multiplication:} As illustrated in Fig.~\ref{fig:photonic_operations}, photonic multiplication can be implemented using two cascaded MZMs. For example, the multiplication of $a_1$ and $b_1$ is conducted using the top two MZMs in the figure. The value $a_1$ is first converted to an RF signal using $DAC_{11}$ to drive the first MZM, which modulates the carrier signal by adjusting its intensity $I_{in}$ as follows: 
\begin{equation}
    I_1 = I_{\text{in}} \cdot f(V_1) = I_{\text{in}} \cdot \cos^2\left(\frac{\pi V_1}{2V_\pi}\right),
\end{equation}
where $f(V_1) = \cos^2\left(\frac{\pi V_1}{2V_\pi}\right)$ is the intensity transfer function of the MZM, $V_1$ is the voltage of the RF signal generated by $DAC_{11}$, and $V_{\pi}$ is the half-wave voltage required to induce a $\pi$ phase shift between the two arms. The output of the first MZM then serves as the input carrier for the second MZM, in which the intensity is further adjusted based on the RF signal representing $b_1$ as follows: 
\begin{equation}
    I_2 = I_1 \cdot f(V_2) = I_{\text{in}} \cdot f(V_1) \cdot f(V_2).
\end{equation}
When both MZMs operate in their linear regions, $f(V_1) \propto V_1$ and $f(V_2) \propto V_2$, leading to $I_{2} \propto V_1 \times V_2$. Therefore, the final output intensity is proportional to the product $a_1 \times b_1$.

\textbf{Photonic Addition:} Photonic addition is achieved by combining optical signals at different wavelengths through a WDM multiplexer, producing an output signal with total optical power corresponding to the sum of the individual inputs~\cite{ishio2003review}. As illustrated in Fig.~\ref{fig:photonic_operations}, the upper and lower pairs of MZMs operate at different wavelengths. When their outputs are merged by the WDM MUX, their optical powers simply accumulate since they do not interfere. Hence, the output of the WDM MUX is $I_{\lambda_1}+I_{\lambda_2}$, which is proportional to $\sum_{i=1}^{2}a_i\times b_i$ (vector dot-product).

\section{Experimental Insights and Motivation for LightMat-HP}\label{sec:exp_mov}

\begin{figure*}[ht]
  \centering
  \begin{subfigure}[t]{0.48\textwidth}
    \centering
    \includegraphics[width=\textwidth]{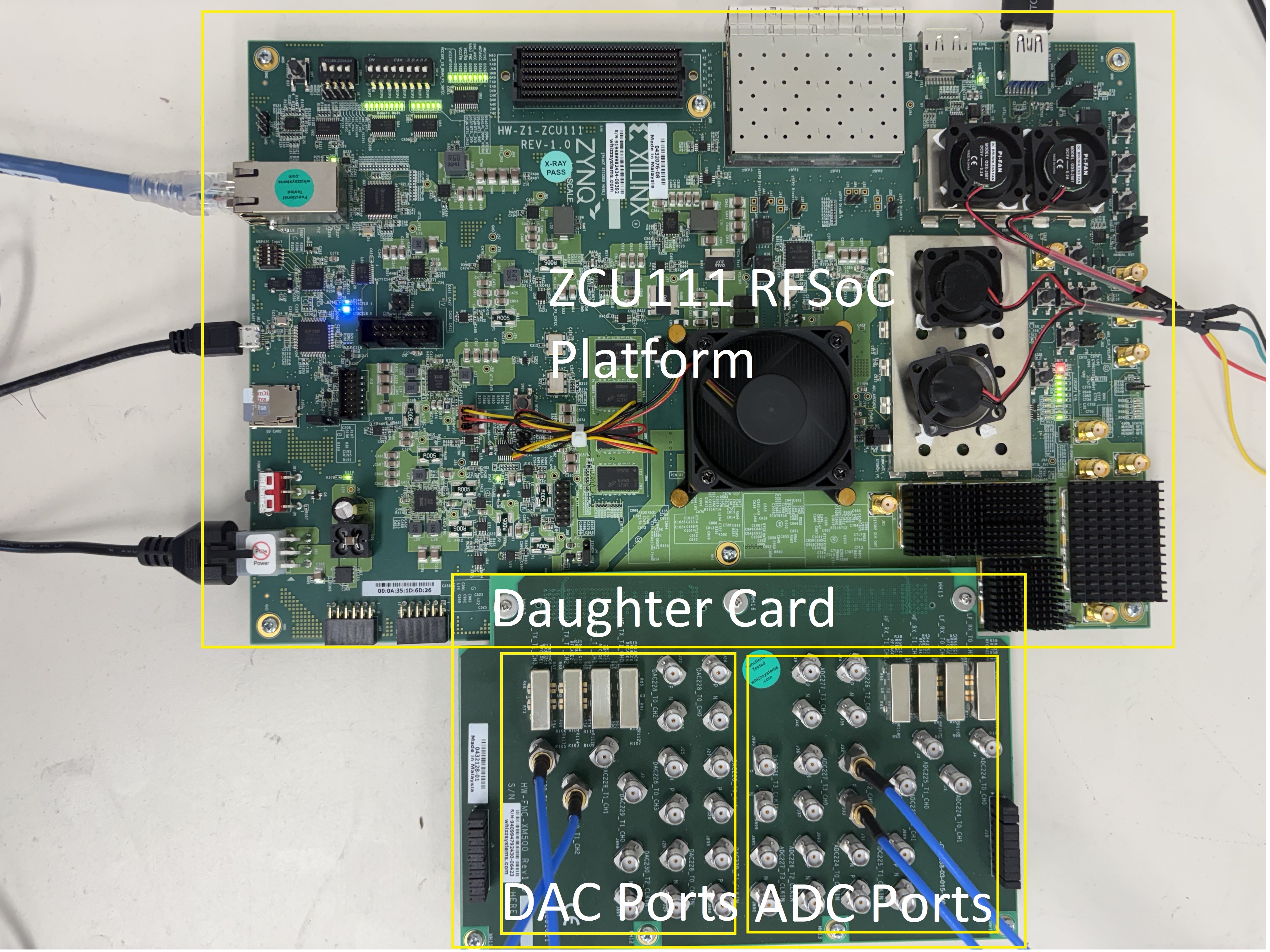}
    \caption{Digital subsystem implemented on the AMD Zynq UltraScale+ RFSoC ZCU111 evaluation board.}
    \Description{}
    \label{fig:digital_component}
  \end{subfigure}
  \begin{subfigure}[t]{0.48\textwidth}
    \centering
    \includegraphics[width=\textwidth]{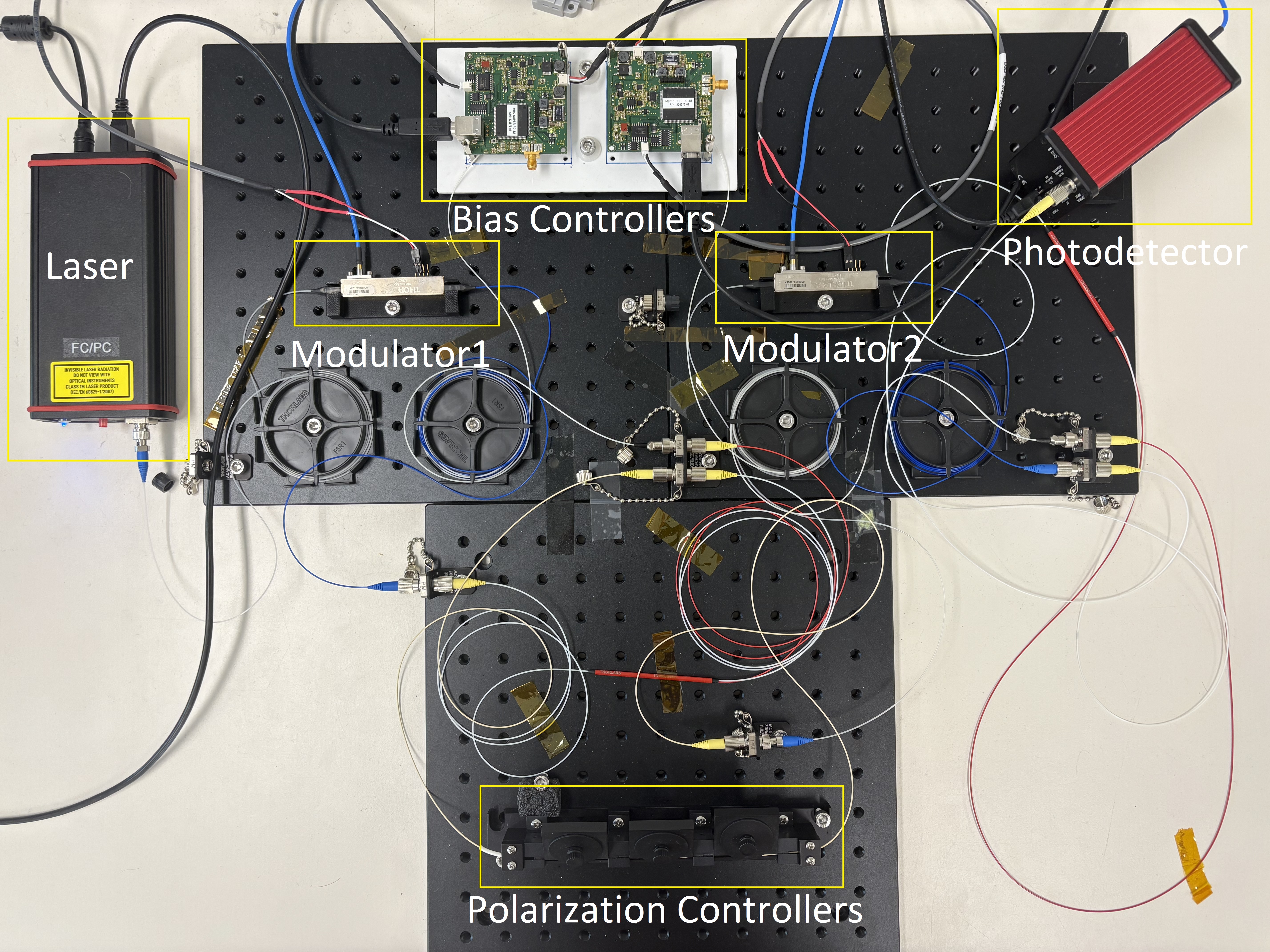}
    \caption{Photonic subsystem consisting of laser, MZMs, bias controller, polarization controller and photodetector.}
    \label{fig:photonic_component}
    \Description{}
  \end{subfigure}
  \caption{The electronic and photonic hardware for implementing the photonic processing prototype. }
  \label{fig:hardware_setup}
\end{figure*}

This section presents an experimental study that provides empirical support and motivations for the design of LightMat-HP. Our experiments were conducted on a photonic computing prototype shown in Fig.~\ref{fig:hardware_setup}. All digital processing was implemented on the Xilinx Zynq UltraScale+ RFSoC ZCU111 evaluation board, which integrates high-end DACs and ADCs capable of sampling at 4 GS/s. We used commercial photonic devices to implement photonic multiplication. A tunable continuous-wave laser~\cite{laserdiodesourceTunableDiode} is used to generate a carrier waveform at $\lambda = 1550 nm$. Two Thorlabs 10 GHz intensity modulators~\cite{thorlabs_lna2322} are used to encode electrical signals onto light.
A Thorlabs polarization controller is placed between the two modulators to maintain the correct polarization state. Two MBC-SUPER bias controllers~\cite{ozoptics_dts0165} are used to lock the modulators at the null point of the intensity transfer function by monitoring the optical output and adjusting the bias voltage. The modulated light is converted to an electrical signal using a Thorlabs RXM15EF photodetector~\cite{thorlabs_rxm15ef}, which has a bandwidth of 15 GHz. 

In practical photonic systems, the numerical accuracy of photonic multiplication is impacted by multiple factors, including modulator nonlinearity and bias drift, optical and electronic noise, and DAC/ADC quantization resolution. All DACs on the ZCU111 feature 14-bit quantization resolution, while all ADCs provide 12-bit quantization resolution, resulting in low DAC/ADC quantization error. In our experiments, we evaluate the cumulative impact of all other factors on computational accuracy. Specifically, we evaluate the accuracy of photonic multiplication across different operand bit-widths by comparing the photonic computing results with the corresponding ground-truth values and analyzing the resulting error characteristics and trends. 
A pair of inputs $(x_1, x_2)$ is fed to the DACs to be  multiplied through the photonic modulators. The output $y_{\text{measured}}$ is compared with the ground truth $y_{\text{ground\_truth}} = x_1 x_2$ using Absolute Error $AE = |y_{\text{measured}} - y_{\text{ground\_truth}}|$ and Relative Error $RE = |y_{\text{measured}} - y_{\text{ground\_truth}}|\times 100/y_{\text{ground\_truth}}$. In our experiments, the bit-width for the operands was varied from $d=3$ to $d=8$. For each bit-width $d$, $x_1$ and $x_2$ were randomly generated within the range $[1, 2^d - 1]$, and 200 integer multiplications were performed to obtain the mean and standard deviation. 

\begin{figure}[!ht]
  \centering

  \begin{subfigure}[t]{0.9\textwidth}
    \centering
    \includegraphics[width=0.85\textwidth]{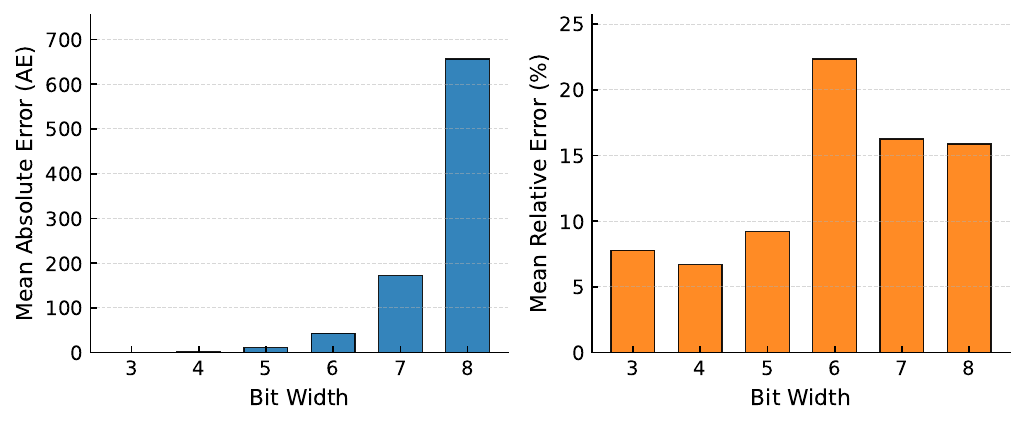}
    \caption{Average absolute error (AE) and relative error (RE) under different quantization bit widths.}
    \label{fig:exp1_AE_RE}
  \end{subfigure}
  \vspace{0.8em}



  \begin{subfigure}[t]{0.40\textwidth}
    \centering
    \includegraphics[width=\textwidth]{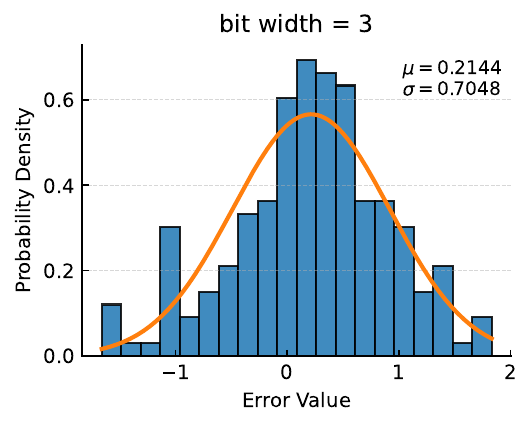}
    \caption{Quantization bit width of 3.}
    \label{fig:exp1_gaussian_dist_bw3}
  \end{subfigure}
  \hspace{0.0\textwidth}
  \begin{subfigure}[t]{0.40\textwidth}
    \centering
    \includegraphics[width=\textwidth]{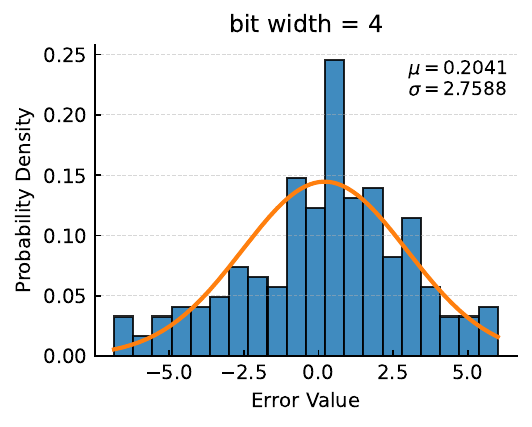}
    \caption{Quantization bit width of 4.}
    \label{fig:exp1_gaussian_dist_bw4}
  \end{subfigure}

  \begin{subfigure}[t]{0.40\textwidth}
    \centering
    \includegraphics[width=\textwidth]{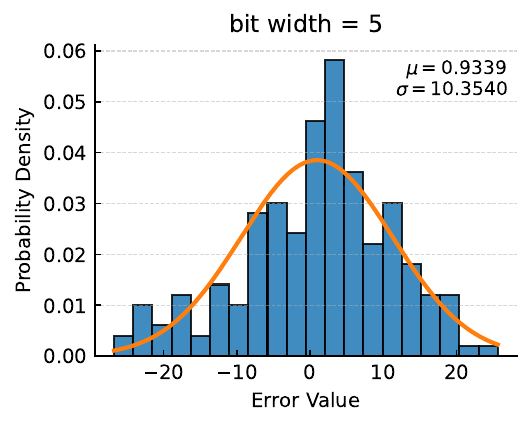}
    \caption{Quantization bit width of 5.}
    \label{fig:exp1_gaussian_dist_bw5}
  \end{subfigure}
  \hspace{0.0\textwidth}
  \begin{subfigure}[t]{0.40\textwidth}
    \centering
    \includegraphics[width=\textwidth]{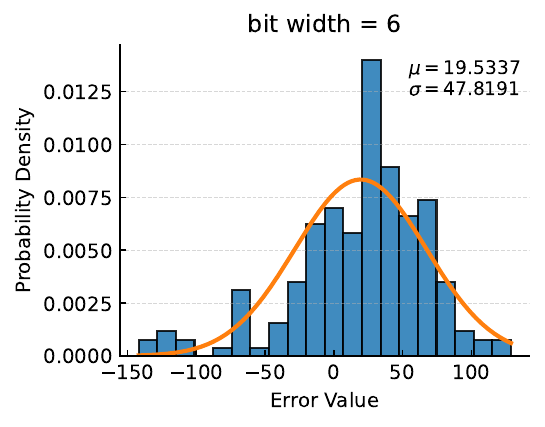}
    \caption{Quantization bit width of 6.}
    \label{fig:exp1_gaussian_dist_bw6}
  \end{subfigure}

  \caption{Error distributions under Gaussian input for different quantization bit widths.}
  \label{fig:exp1_gaussian_dist_all}
\end{figure}

Fig.~\ref{fig:exp1_gaussian_dist_all} (a) shows that, \textbf{as the bit width increases, the absolute error (AE) grows rapidly, whereas the relative error (RE) demonstrates a generally increasing but non-monotonic trend}. At low bit-width ($d=3$), the system reports an AE of $0.61$ and an RE of $7.79\%$. When the bit width increases to $d=7$, AE increases to $171.98$, with an RE of $16.26\%$. This trend suggests that photonic multiplication is highly accurate at low bit-widths but becomes increasingly sensitive to noise and analog imperfections as the operand dynamic range expands. 
This is because that computational accuracy is fundamentally limited by the system signal-to-noise-and-distortion ratio (SINAD) rather than by quantization resolution alone. Achieving $n$-bit precision requires $2^n$ distinguishable ADC readouts. Due to the limited linearity of photonic modulators and the analog noise and distortion, the effective number of bits (ENOB), determined by the measured SINAD, imposes an upper bound on the precision of the photonic multiplication system. When the bit width exceeds this SINAD-limited ENOB, additional bits primarily represent noise and distortion components instead of meaningful signal information. Consequently, noise and nonlinearity introduced by the photonic multiplier, including electro-optic modulator nonlinearity, laser relative intensity noise, shot noise, and distortion from mixed-signal electronic interfaces, dominate the error behavior. As bit width increases, these impairments accumulate across more significant bits, resulting in a monotonic increase in both absolute and relative errors. The observed trend confirms that the system is not quantization limited, but rather constrained by physical noise and distortion mechanisms. Therefore, increasing numerical precision beyond the SINAD-supported resolution does not improve, and can even degrade, computational accuracy.

Fig~\ref{fig:exp1_gaussian_dist_all} (b)-(e) further show that the error distributions approximately follow Gaussian distributions, suggesting that the errors are dominated by stochastic noise rather than systematic bias. Increasing the bit-width leads to higher mean and variance in the error distribution. This behavior arises from the combined effects of bit-width configuration and photonic–electronic system noise. At low bit widths, coarse quantization steps partially mask device nonlinearity and noise, resulting in smaller normalized errors. As the bit width increases, the photonic–electronic chain becomes increasingly sensitive to drift, noise, and transfer-function mismatch, which do not scale down with the quantization step size. Consequently, higher bit-widths lead to an increase in the mean absolute error.

To measure the accuracy of dot-product computation on our testbed, we evaluate the relative error between BFP-based photonic dot products and electronic FP-based dot products, with different bits sliced mantissa multiplication. The input vector elements follow a uniform distribution $\mathcal{U}[1,100]$ and are encoded using a BFP representation with a shared 8-bit exponent. During the computation of the dot product, the photonic multiplication of mantissa is implemented through a decomposition of $4$-bit $\times$ $4$-bit (blue line) and $5$-bit $\times$ $5$-bit (orange line). The mantissa slice method is illustrated in Fig~\ref{fig:mantissa_multiplication} (a), and the dot product is obtained by accumulating the decomposed mantissa products and rescaling by the combined exponent $E_x+E_y$ in the digital domain. Each partial product is inevitably affected by noise introduced by the photonic computing units and by DAC/ADC interfaces. We evaluate dot products with vector length $K$ ranging from $32$ to $1024$, repeating each configuration for $1000$ trials. The FP32 results obtained from the software are considered the ground truth for measuring the relative error (RE). The results show that the mean relative error remains stable as $K$ increases, while the error dispersion is progressively reduced.
Since the error distribution for low-bit photonic multiplication has an approximately zero-mean Gaussian distribution, the computational errors could be canceled out to some extent during digital accumulation, leading to stable and predictable numerical behavior for larger vector lengths. With 5-bit slicing, the relative error (RE) remains below 0.75\%. With 4-bit slicing, the relative error further decreases and remains below 0.7\% for large vectors. This demonstrates that BFP-based dot-product computation on a real testbed achieves controlled and scalable numerical accuracy, which directly motivates the adoption of BFP arithmetic in our architecture for precision-configurable photonic matrix multiplication.

\begin{figure}[thbp]
  \includegraphics[width=0.9\textwidth]{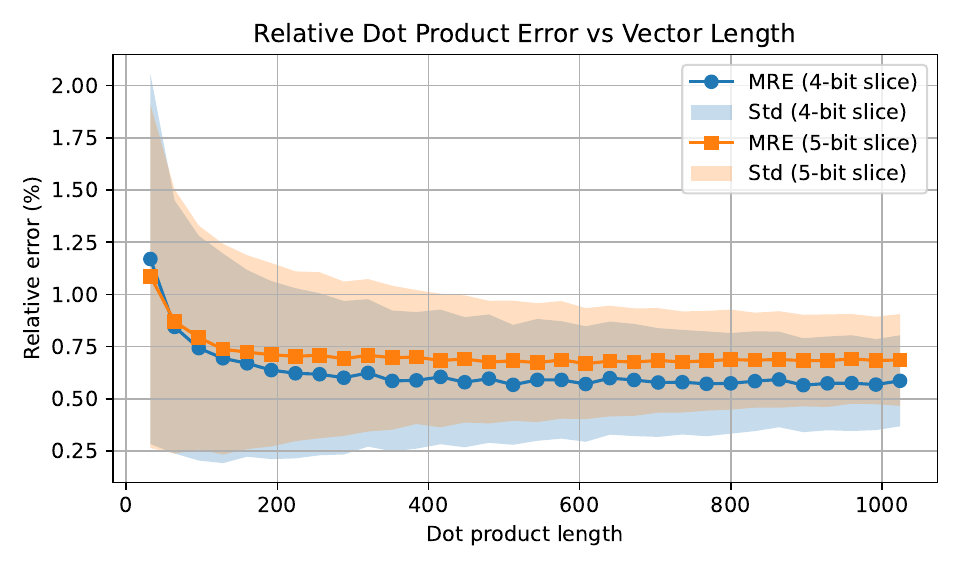}
  \caption{Dot-product accuracy comparison between FP32 and BFP representations under photonic noise.}\label{fig:error_dot}
  \Description{}
\end{figure}

The above experimental findings provide the \textbf{design motivations} for the proposed LightMat-HP system:

\begin{itemize}
    \item \textbf{Given the limited precision of photonic multiplication (3-6 bits), high bit-width arithmetic is more effectively realized through decomposition into low bit-width arithmetic rather than relying on higher analog precision.}
    Our results indicate that accurate high bit-width multiplication can be constructed by decomposing operands into low-bit-width components that match the reliable operating range of the photonic hardware, followed by digital accumulation and reconstruction.
    
    \item \textbf{The testbed measurements demonstrate that BFP-based dot-product computation with decomposed mantissa multiplication achieves stable and predictable accuracy.} This observation, together with the advantages of block floating-point (BFP) arithmetic in reducing computational complexity and memory bandwidth requirements while enabling a flexible tradeoff between precision and performance, motivates the adoption of BFP as the core numerical format in LightMat-HP.
\end{itemize}

\begin{figure}[thbp]
  \includegraphics[width=0.9\textwidth]{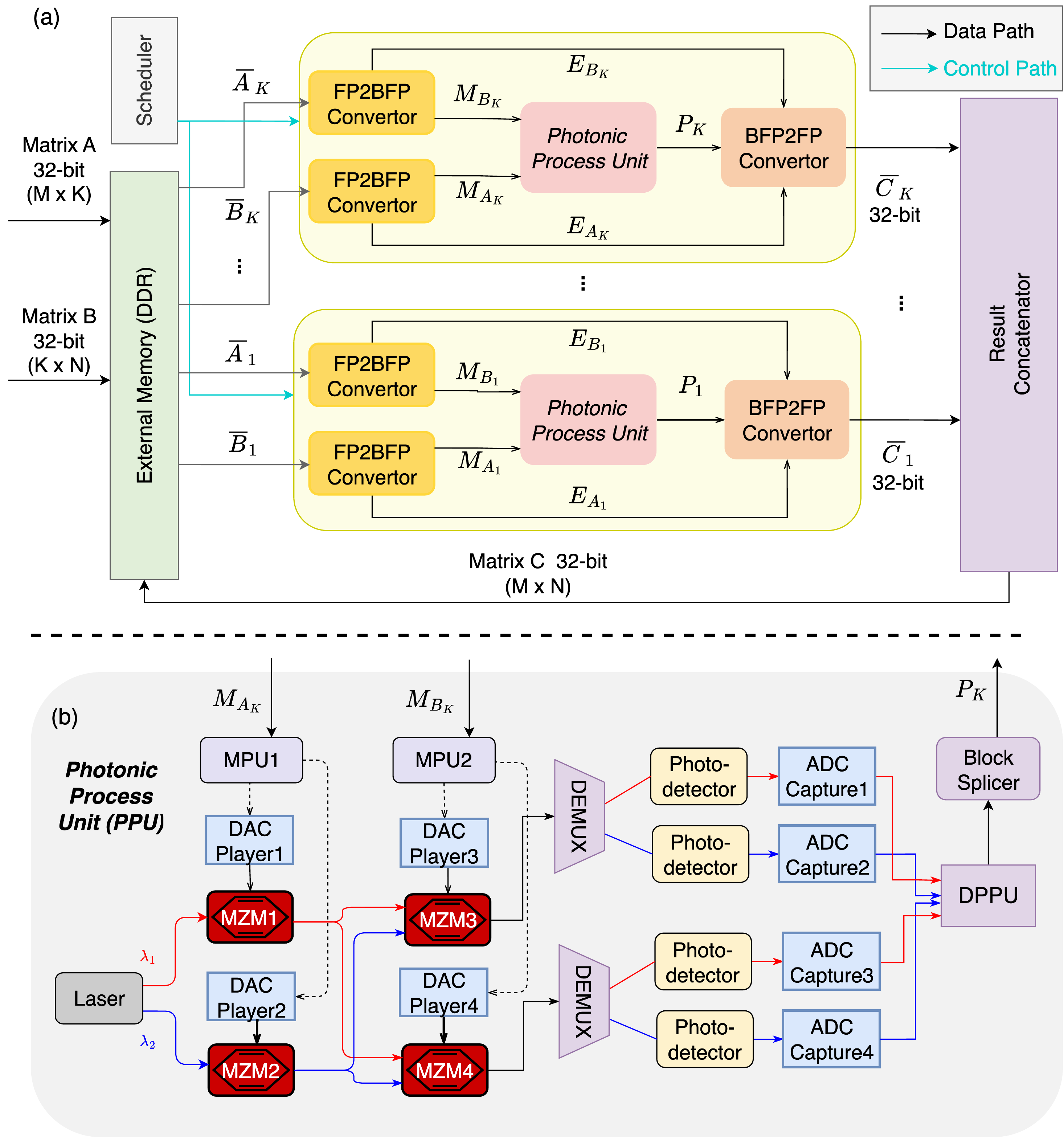}
  \caption{(a) Architecture of the LightMat-HP system. (b) Illustration of the internal structure of Photonic Processing Unit.}\label{fig:sys_architecture}
  \Description{}
\end{figure}

\section{LightMat-HP: System Architecture and Design}\label{sec:LightMat-HP}

\subsection{LighMat-HP Architecture}
\subsubsection{Architecture Overview} LightMat-HP is a hybrid photonic–electronic computing system designed to accelerate matrix multiplication on a wide range of scales with configurable computational precision. By combining the advantages of electronic control and storage with the high speed and low power consumption of photonic computation, LightMat-HP offers an efficient platform for high-performance matrix processing and provides insights for the design of future general-purpose hybrid photonic–electronic computing platforms.

As shown in Fig.~\ref{fig:sys_architecture}, LightMat-HP consists of a digital subsystem and a photonic subsystem.
The digital subsystem includes external memory, a scheduler, two precision conversion modules (FP2BFP and BFP2FP) for the conversion between FP and BFP, and a result concatenator for assembling matrix outputs. The photonic subsystem comprises the components required to perform high-speed multiplication in the optical domain, together with the necessary digital interfaces. Fig.~\ref{fig:sys_architecture} (b) illustrates a simplified photonic multiplication core, which includes a laser source, four MZMs, and four photodetectors.
Optical signal reuse is achieved through optical splitting and interconnection to enable parallel analog multiplication within a single Photonic Processing Unit (PPU).
The architecture can support higher throughput by adding more PPUs operating in parallel.

\subsubsection{End-to-End Dataflow for Matrix Multiplication}
The end-to-end execution flow of matrix multiplication in LightMat-HP is described as below.
Input matrices $\mathbf{A}$ and $\mathbf{B}$ are stored in external memory in the original FP format (e.g., FP16, FP32 and FP64).
The electronic scheduler partitions the matrices into computation tiles (i.e., submatrices) according to a predefined scheduling policy and streams them to the corresponding PPUs.

Before entering the photonic domain, the FP operands are converted to BFP representation. 
For the $k$-th matrix tile, the shared exponents $E_{A_k}$ and $E_{B_k}$ are forwarded directly to the BFP2FP converter for accumulation in the digital domain due to its simplicity, while the mantissas $M_{A_k}$ and $M_{B_k}$ are streamed into the PPU for multiplication in the optical domain. Each photonic processing unit generates partial products for the corresponding matrix multiplication in parallel, exploiting the bandwidth and parallelism of analog photonic computation.

The photonic output produces the partial result $P_k$, which is digitized and converted back to the original FP format in the electronic domain.
During this process, normalization and accumulation are performed using digital arithmetic.
The result concatenator finally combines partial results from multiple PPUs to form the output matrix $\mathbf{C}$ in the original FP format.

Although addition operations can be performed in the optical domain, LightMat-HP adopts digital addition rather than photonic addition for accumulating exponents, slice-level products, and element-level products due to the following reasons. First, addition is significantly less expensive than multiplication and can be implemented more efficiently in the digital domain. Second, digital accumulation avoids analog noise accumulation and scales robustly with accumulation depth, which is critical for achieving high accuracy in matrix multiplication.

\subsubsection{Photonic Processing Unit} The PPU in LightMat-HP is designed to perform high-throughput analog multiplication for BFP mantissas. As illustrated by Fig.~\ref{fig:sys_architecture} (b), the PPU accepts two input data streams corresponding to the mantissas of two matrix tiles $M_{A_k}$ and $M_{B_k}$. These mantissas are first processed by the Mantissa Processing Unit (MPU), where each mantissa is decomposed into multiple low bit-width slices according to the bit-width of the mantissa. These slices are then converted to parallel digital streams and fed into the corresponding \textit{DAC players} modules, which drive the associated DACs at a fixed sampling rate to generate the corresponding analog signals used for encoding digital slices onto optical carriers through MZMs.

The PPU employs four MZMs to perform four parallel photonic multiplications per cycle, with each pair of multiplications sharing a common operand. To be specific, two wavelength carriers $\lambda_1$ and $\lambda_2$ are generated by the laser source and independently modulated by the first two MZMs ($MZM_1$ and $MZM_2$) to encode two common operands. The outputs of $MZM_1$ and $MZM_2$ are both equally split and routed to the remaining two MZMs, enabling parallel multiplications with distinct secondary operands. The resulting optical signals are then demultiplexed by wavelength, detected by independent photodetectors, and converted into digital samples by ADCs. The ADC outputs are buffered at the \textit{ADC Captures} modules and forwarded to the Digital Post-Processing Units (DPPUs), which accumulate and align the products of slice multiplications to generate the result of mantissa multiplication. Finally, the block splicer assembles the multiplication results belonging to the same matrix title $P_k$ and streams them back to the electronic subsystem. As shown in Fig. \ref{fig:sys_architecture} (a), multiple PPUs can operate concurrently, enabling scalable parallel execution without modifying the internal structure of each unit.

\subsection{Tile-Based Precision-Configurable Photonic Matrix Multiplication} 
\subsubsection{BFP-FP-Conversion in Tiled Matrix Multiplication}

To support matrix multiplication of arbitrary sizes, LightMat-HP partitions the input matrices into fixed-size tiles before computation, with each tile serving as the basic unit of computation and scheduling. Building on this structure, LightMat-HP adopts a BFP representation that is tightly coupled with the tiled matrix multiplication, i.e., \textit{BFP blocks are constructed from rows of matrix $A$ and from columns of matrix $B$, respectively}.
The number of rows to form a BFP block in a tile of matrix $A$, as well as the number of columns to form a BFP block in a tile of matrix $B$ are both configurable parameters.


The FP2BFP and BFP2FP modules are the components in LightMat-HP for converting numerical values between the FP and BFP representations. The FP2BFP converter performs the block-wise conversion from FP to BFP, as explained in Section \ref{sec:bfp}. The number of mantissa bits is a configurable system parameter, enabling exploration of the tradeoff between computational precision and complexity. For example, if the input matrices are in FP32 format, they can be converted to BFP with a mantissa bit width of 20 to preserve computational accuracy, or with a mantissa bit width of 10 to reduce computational complexity at the cost of reduced accuracy. After photonic computation, the BFP2FP module reconstructs the results from the block-level BFP representation to the original FP format. For each element, the output value is recovered as $\hat{x}_i = m_{b_i} \cdot 2^{e_s}$, followed by standard floating-point re-encoding. Since exponent recovery corresponds to a deterministic scaling operation, BFP2FP itself does not introduce additional system noise. 


\subsubsection{Precision-Configurable Mantissa Multiplication via Slice-Based Decomposition}\label{sec:slice_decomp}

\begin{figure}[th!]
  \includegraphics[width=\textwidth]{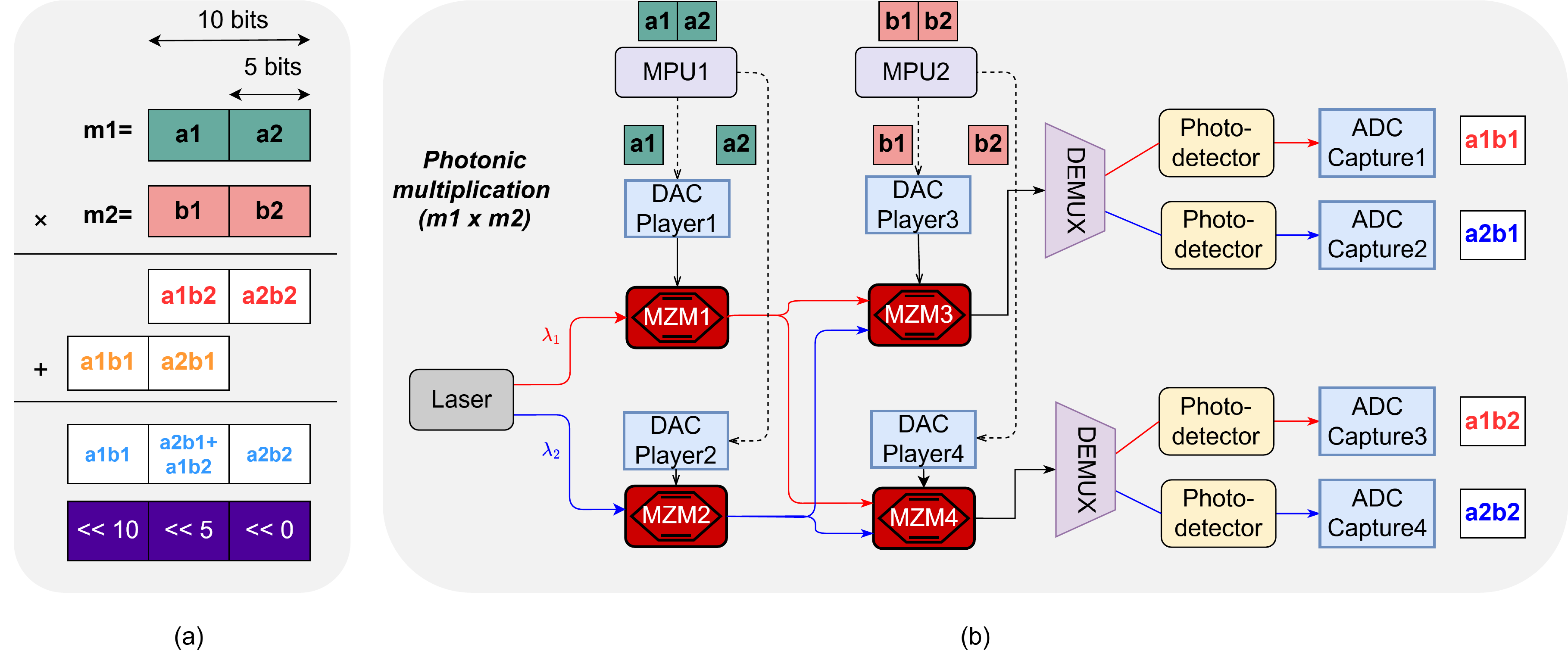}
  \caption{(a) Illustration of the multiplication process of two 10-bit mantissas. (b) Mapping 5-bit mantissa blocks to the LightMat-HP photonic computing unit.}\label{fig:mantissa_multiplication}
  \Description{}
\end{figure}

Fig.~\ref{fig:mantissa_multiplication} illustrates the proposed precision-configurable mantissa multiplication scheme based on slicing. Motivated by the experimental study and the analysis of computational error in Section~\ref{sec:exp_mov}, the key idea of this scheme is to decompose a high bit-width mantissa multiplication into multiple low bit-width sub-multiplications. The slice-based decomposition process for mantissas is implemented in the MPU as shown in Fig.~\ref{fig:matrix-multiplication}. Fig.~\ref{fig:mantissa_multiplication} (a) illustrates the multiplication of 10-bit mantissas, where both $m_1$ and $m_2$ are decomposed into two 5-bit slices, denoted as $(a_1, a_2)$ and $(b_1, b_2)$, respectively. The original 10-bit multiplication is then expanded into four 5-bit-by-5-bit sub-products: $a_1b_1$, $a_2b_1$, $a_1b_2$, and $a_2b_2$. These sub-products are then combined through appropriate shifting and accumulation to reconstruct the full-precision multiplication result. Figure~\ref{fig:mantissa_multiplication} (b) shows how the sub-multiplications are mapped to the PPU computing pipeline. The slices $a_1$ and $a_2$ are serialized and applied to MZM1 and MZM2, respectively; similarly, slices $b_1$ and $b_2$ are applied to MZM3 and MZM4, respectively. 
Through optical splitting and wavelength division multiplexing/de-multiplexing, the optical signal output from MZM3 represents the sub-products $a_1b_1$ with $\lambda_1$ and $a_2b_1$ with $\lambda_2$, while the output from MZM4 represents the sub-products $a_1b_2$ with $\lambda_1$ and $a_2b_2$ with $\lambda_2$. This design enables us to compute all partial products within one compute cycle. Finally, the modulated optical signals are detected by photodetectors and captured by ADCs, producing digitized sub-product results. In the subsequent electronic processing stage, these sub-products are shifted according to their corresponding bit positions and accumulated to reconstruct the final mantissa multiplication result in the DPPU module. 

If each mantissa is decomposed into more than two slices, the resulting sub-multiplications can either be mapped to multiple PPUs for parallel computation in a single cycle or mapped to a single PPU over multiple cycles. For example, the multiplication of two 20-bit mantissas with each divided into four 5-bit slices yields a total of 16 sub-multiplications. These sub-multiplications may be distributed across four PPUs, allowing all 16 sub-multiplications to be completed in a single cycle. Alternatively, they may be mapped to a single PPU, in which case four cycles are required to complete all sub-multiplications, since each PPU can perform only four parallel multiplications per cycle.

This method enables scalable, high-precision multiplication in photonic computing units with limited analog precision. Through mantissa slicing, the overall multiplication accuracy is no longer restricted by the analog precision of a single photonic multiplication, instead by the selection of slice bit-width, the sub-multiplications in the photonic domain, and the subsequent accumulation performed in the digital domain. Moreover, this method supports run-time configurability of mantissa multiplication precision. On one hand, the slice bit-width can be selected according to the experimentally observed accuracy characteristics of the photonic computing unit at different precisions, ensuring that all photonic sub-multiplications operate within a reliable bit-width range. On the other hand, higher numerical precision can be achieved by increasing the number of slices, trading latency for precision without requiring high-cost photonic devices for improving analog computation precision. As a result, by leveraging the advantages of high-throughput and parallelism in photonic computing, this approach provides a controllable and scalable solution for precision-configurable photonic computing.

\subsubsection{Tile-based Matrix Multiplication}\label{sec:tile_matrix_multiplication}
\begin{figure}[th!]
\includegraphics[width=0.8\textwidth]{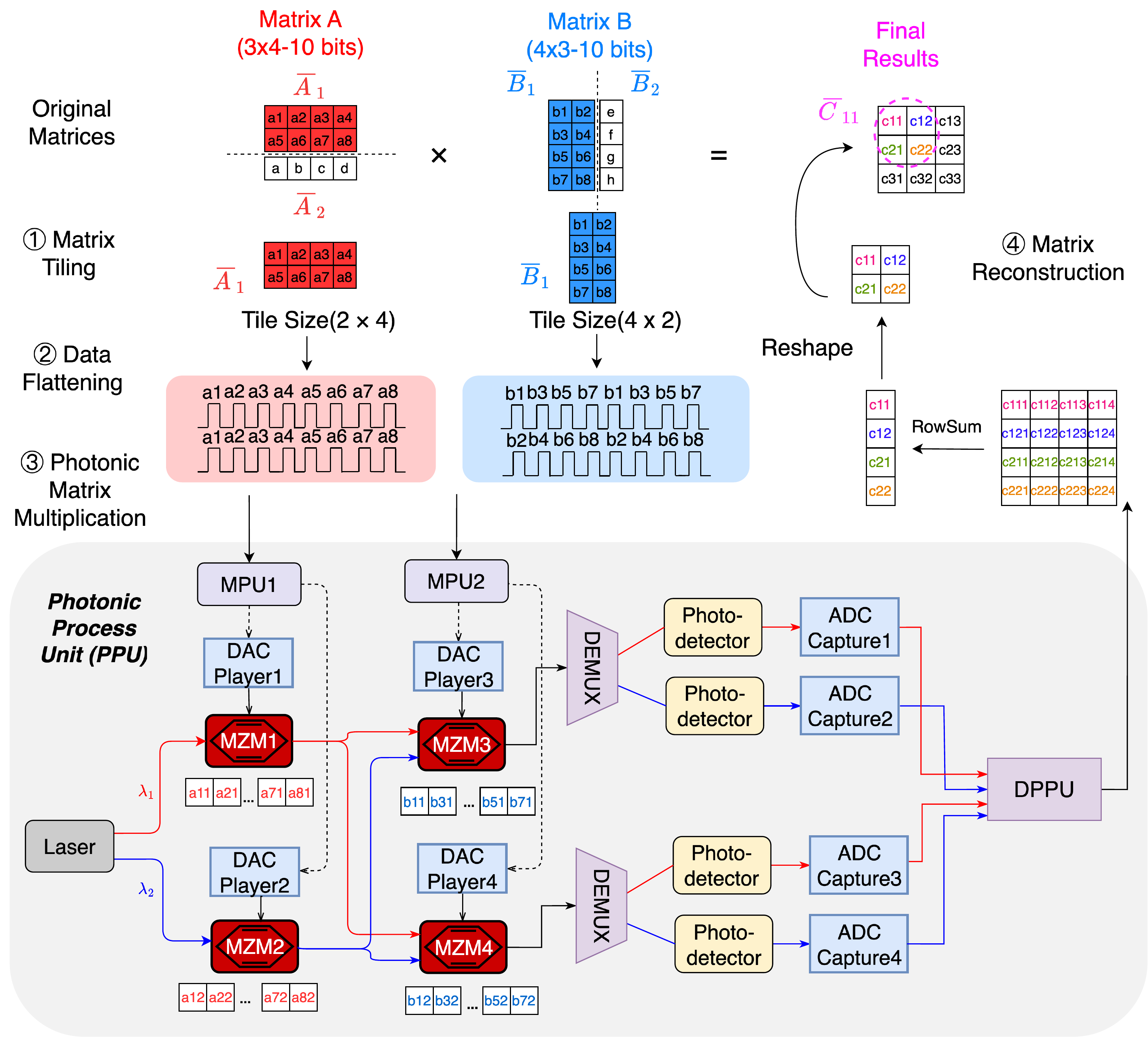}
  \caption{Illustration of the tile-based photonic matrix multiplication, including matrix tiling, data flattening, optical-domain multiplication, 
  and digital reconstruction of the final results.}\label{fig:matrix-multiplication}
  \Description{}
\end{figure}

To support matrix multiplication of arbitrary sizes, LightMat-HP uses a tile-based matrix multiplication dataflow in which a large matrix is partitioned into fixed-size tiles. Typically, the tile size is defined as $L \times K$, where $L$ and $K$ represent the number of rows and columns in the tile, respectively. For $\mathbf{A} \times \mathbf{B}$, $\mathbf{A}$ is partitioned along the row dimension into tiles $\{\bar{A}_{i}\}$ and $\mathbf{B}$ is tiled along the column dimension into tiles $\{\bar{B}_{i}\}$. Hence, all tiles in $\mathbf{A}$ have the same number of columns as $\mathbf{A}$, and all tiles in $\mathbf{B}$ have the same number of rows as $\mathbf{B}$. 
For the BFP representation, each row of the tile matrix $\bar{A}_{i}$ is treated as an independent BFP block sharing a common exponent, while each column of the tile matrix $\bar{B}_{i}$ is represented as a separate BFP block with its own shared exponent. 
The tile size $L \times K$ is determined by the size of SRAMs in the \textit{DAC Players} (32KB in LightMat-HP), ensuring that the operands for all multiplications of a tile pair can be loaded into the SRAMs of the two \textit{DAC Players} so that the multiplication of a tile pair can be executed with a single trigger. The whole process of tiled photonic matrix multiplication on the mantissa blocks is illustrated in Fig.~\ref{fig:matrix-multiplication}, which consists of four steps: \textit{matrix tiling}, \textit{data flattening}, \textit{photonic matrix multiplication}, and \textit{matrix reconstruction}. The pseudocode for tile-based matrix multiplication is given in Algorithm \ref{alg:tiled_photonic_matmul}. 

\begin{algorithm}[htbp]
\caption{MatrixMultiply(A, B)}
\label{alg:tiled_photonic_matmul}
\KwIn{$A\in\mathbb{R}^{M\times K}$, $B\in\mathbb{R}^{K\times N}$: input matrices; $[L,K], [K,L]$: dimensions of tile.}
\KwOut{$C$: result matrix}

\textbf{Step 1: Matrix Tiling:} \\
\tcc{Decompose $A$ into $L {\times}K$ tiles and , $B$ into $K{\times} L$ tiles}
$
A = [\bar{A}_{i}],\quad B = [\bar{B}_{j}], \quad 1\le\!i\le\!M/L, 1\le\!j\le\!N/L.
$

\textbf{Step 2: Data Flattening:} \\
\ForEach{tile pair $(\bar{A}_{i}, \bar{B}_{j})$}{
$
V_1 = \mathrm{Flatten}_r(\bar{A}_{i}, K, L)$;\\
$V_2 = \mathrm{Flatten}_c(\bar{B}_{j}, K, L)$; \\

Append $(V_1,V_2)$ to the transmission buffer $\mathcal{B}$\;
}


\textbf{Step 3: Photonic Multiplication:} \\
\tcc{Using PPU to perform photonic multiplication}
\ForEach{voltage pair $(V_1, V_2)$ in $\mathcal{B}$}{
    $res_{1}, res_{2}, res_{3}, res_{4} \leftarrow \mathrm{PhotonicMultiply}(V_1,\, V_2)$ \\
    \tcc{$res_{i}$ corresponding to the result of each ADC Capture}
    Append $res_{i}$ to result buffer $\mathcal{R}_{i}$\ \\
}

\textbf{Step 4: Matrix Reconstruction:} \\
\tcc{Aggregate tile multiplications}
$C = \mathrm{MatrixReconstruct}(\mathcal{R}_{1},\mathcal{R}_{2}, \mathcal{R}_{3},\mathcal{R}_{4}, L, K, M, N)$ \\
\Return{$C$}\;
\end{algorithm}

\textbf{Matrix Tiling:} Given the mantissa matrices $A\in\mathbb{R}^{M\times K}$, $B\in\mathbb{R}^{K\times N}$ with the output $C=A B\in\mathbb{R}^{M\times N}$, and a tile size $L\times K$, we know that the matrices $A$ and $B$ can be decomposed into $\hat{M}$ and $\hat{N}$ tiles respectively, where:
\begin{equation}
\hat{M} = \left\lceil \dfrac{M}{L} \right\rceil L, 
\hat{N} = \left\lceil \dfrac{N}{L} \right\rceil L, 
\label{eq:padding_dims}
\end{equation}
We tile $A$, $B$, and the output $C=AB\in\mathbb{R}^{M\times N}$ into $L\times K$ blocks except the last tile:
\begin{equation}
A=\big[\bar{A}_{p}\big], 
B=\big[\bar{B}_{r}\big],
C=\big[\bar{C}_{p,r}\big]
\end{equation}
where $1\le p \le \hat{M}/L$, $1\le r \le \hat{N}/L$, and each tile $\bar{A}_{p}\in\mathbb{R}^{L\times K}$, $\bar{B}_{r} \in\mathbb{R}^{K\times L}$, $\bar{C}_{p,r}\in\mathbb{R}^{L \times L}$. The last tile of matrix $A$ is formed by the remaining $(\hat{M} \bmod L)$ rows, while the last tile of matrix $B$ is formed by the remaining $(\hat{N} \bmod L)$ columns. Step 1 in Fig. \ref{fig:matrix-multiplication} illustrates the tiling of Matrix A (3 $\times$ 4) and Matrix B (4$\times$ 3).

\textbf{Data Flattening:} Each tile pair is transformed into one-dimensional sequences to match the temporal input format of the DACs. For the tile multiplication $\bar{A}_{i} \times \bar{B}_{j}$, $\bar{A}_{i}$ is flattened in a row-major order, where the matrix is unfolded row by row to form a one-dimensional sequence of length $L \times K$, and then the $L \times K$ sequence is repeated $L$ times. In contrast, $\bar{B}_{j}$ is flattened in a column-major order by unfolding the matrix column by column into a one-dimensional sequence, with each column repeated $L$ times before flattening the next column. As illustrated in Fig.~\ref{fig:matrix-multiplication}, the tile $\bar{A}_{1}$ is flattened into a vector $V_{1} = [a_{1}, a_{2}, a_{3}, a_{4}, a_{5}, a_{6}, a_{7}, a_{8}, a_{1}, a_{2}, a_{3}, a_{4}, a_{5}, a_{6}, a_{7}, a_{8}]$, and the tile $\bar{B}_{1}$ is flattened into a vector $V_{2} = [b_{1}, b_{3}, b_{5}, b_{7}, b_{1}, b_{3}, b_{5}, b_{7}, b_{2}, b_{4}, b_{6}, b_{8}, b_{2}, b_{4}, b_{6}, b_{8}]$. The pair-wise multiplication of $V_{1}$ and $V_{2}$ will compute the results of all multiplications in $\bar{A}_{i} \times \bar{B}_{j}$.  

\textbf{Photonic Multiplication:} The two vectors $V_1$ and $V_2$ are aligned and streamed to MPU1 and MPU2, respectively. Within each MPU, the mantissas are sliced and dispatched to the corresponding \textit{DAC players}. In the example shown in Fig.~\ref{fig:matrix-multiplication}, each $a_i$ is decomposed into two slices, $a_{i1}$ and $a_{i2}$, and each $b_i$ is decomposed into two slices, $b_{i1}$ and $b_{i2}$. Following the slicing method presented in Section \ref{sec:slice_decomp}, all $a_{i1}$ slices are streamed to MZM1 for modulation, while all $a_{i2}$ slices are streamed to MZM2. Similarly, all $b_{i1}$ slices are streamed to MZM3, while all $b_{i2}$ slices are streamed to MZM4. The complementary row-major and column-major flattening scheme, together with the mantissa slicing and WDM-based parallel modulation, naturally aligns row–column element pairs without introducing additional control logic. Such alignment simplifies accumulation, shifting, and data layout in the digital post-processing stage. As a result, the photonic computing core efficiently computes the partial products required for tile-based matrix multiplication.

\begin{algorithm}[htbp]
\caption{MatrixReconstruct$(\mathcal{R}_{1},\mathcal{R}_{2}, \mathcal{R}_{3},\mathcal{R}_{4}, L, K, M, N)$}
\label{alg:matrix_reconstruct}
\KwIn{$\mathcal{R}_{k}$: pair-wise multiplication results; $[L,K], [K,L]$: dimensions of tile; $[L, L]$: dimensions of result matrix.}
\KwOut{$C$: reconstructed matrix.}

\tcc{Initialization}
$m \leftarrow \left\lceil M/L \right\rceil, n \leftarrow \left\lceil N/L \right\rceil $ \\
$M[0:m][0:n] = \mathbf{0}$;\\

\textbf{Step 1: Tile-wise reconstruction}\\
\For{$j \leftarrow 1$ \textbf{to} $m$}{
    \For{$j \leftarrow 1$ \textbf{to} $n$}{
        $\mathcal{S} \leftarrow \mathrm{DDPU}
        (\mathcal{R}_{1}, \mathcal{R}_{2}, \mathcal{R}_{3}, \mathcal{R}_{4})$ \\
        $s \leftarrow \mathrm{Organize}(\mathcal{S})$; \\ \label{line:alg_2_reorgan}
        $s \leftarrow \mathrm{RowSum}(s)$; \\
        $M[i][j] \leftarrow \mathrm{Reshape}(s, (L,L))$; \\
    }
}

\textbf{Step 2: Combine sub-matrices}\\
$\bar{C} \leftarrow \mathrm{BlockConcat}(M)$;\\ \label{line:alg2_concat}
\textbf{Step 3: Crop valid region}\\
$C \leftarrow \bar{C}[1{:}M,\ 1{:}N]$;\\ \label{line:alg2_crop}
\Return{$C$};
\end{algorithm}

\textbf{Matrix Reconstruction:} Algorithm~\ref{alg:matrix_reconstruct} gives the pseudocode to reconstruct the result matrix $C$. The pair-wise multiplication results for the slices buffered in \textit{ADC Players} are forwarded to the DPPU, where bit-shifting and accumulation operations are performed to reconstruct the results for high bit-width mantissa multiplications (line 6). 


Let $\mathcal{S}$ denote the data produced by DPPU that contains the high bit-width result of mantissa multiplications. These results are first partitioned into $L$ arrays, each containing $K$ elements. The resulting $\mathbb{N} \times L$ arrays are then reorganized to enable row-wise accumulation (lines 7-8), where $\mathbb{N}$ denotes the level of parallelism inside the PPU and is determined by the PPU architecture. Next, the $K$ elements in each row are summed to generate $L \times L$ results. Finally, these values are reshaped into a result tile of size $L \times L$ (line 9). After all tiles are reconstructed, they are concatenated to form the final result \(C\) (line~\ref{line:alg2_concat}). 

Our design on tile-based matrix multiplication has several advantages: (a) resource constraints are eliminated because each photonic multiplication operates on fixed-size tiles that fit memory limits; (b) it enables pipeline scheduling because the tile-based structure creates a predictable and uniform data flow; (c) scalability is inherent because additional photonic units or DAC/ADC channels can be added to process tiles in parallel.

\section{Precision Modeling and Error Analysis of LighMat-HP}\label{sec:precision_analysis}

This section models the numerical precision and performs error analysis of the proposed LightMat-HP system.
As shown in Fig.~\ref{fig:sys_architecture}, each PPU processes one pair of matrix tiles. For the $k$-th pair of input data, the input matrices are represented as $\mathbf{A}_k = 2^{E_{A,k}} \mathbf{M}_{A,k}, \mathbf{B}_k = 2^{E_{B,k}} \mathbf{M}_{B,k}$, where $E_{A,k}$ and $E_{B,k}$ are shared exponents, and $\mathbf{M}_{A,k}$ and $\mathbf{M}_{B,k}$ are mantissa matrices. The PPU performs photonic multiplication on the mantissas, while exponent recovery and result accumulation are carried out in the digital domain by the BFP2FP converter.

In this computation model, numerical errors in LightMat-HP originate from three main sources: \textit{(1) quantization errors introduced during conversion from FP to BFP format, (2) system error of the PPU arising from photonic multiplication, and (3) rounding errors introduced during digital accumulation and FP formatting.} 


\subsection{FP2BFP Quantization Error.} For the first stage, during the conversion from FP to BFP representation, all elements within the same block share a common exponent $E_k$.
Let $b$ denote the mantissa bit width, $K$ denote the number of block, and $N$ denote the number of values in a block. For any element $x_i$ within a block, the absolute quantization error introduced by BFP representation satisfies
\begin{equation}\label{eq:quant_error}
    \Delta^{quant}_{i} = | \hat{x}_i - x_i | \le  2^{E_{m}} \cdot 2^{-(b+1)}.
\end{equation}
where $E_{m}$ is the maximum value of the common exponent $E_k$. Therefore, for a tile with $N \times K$ numbers, the total error is bounded by:

\begin{equation}\label{eq:quant_error}
    \Delta^{quant}  \le  N \cdot K \cdot 2^{E_{m}} \cdot 2^{-(b+1)}.
\end{equation}
This bound indicates that quantization error under BFP representation is determined by the shared exponent and mantissa precision, rather than by element-wise exponent variation. When extended to a matrix tile containing $N \times K$ elements, it remains predictable and bounded.

\subsection{Photonic Multiplication Error}



Experimental results in Section~\ref{sec:exp_mov} indicate that, the deviation between photonic multiplication outputs and ideal digital results approximately follows a Gaussian distribution. 
Based on these observations, we adopt a Gaussian model for noise modeling. We associate the noise with each photonic multiplication, which more accurately reflects the physical computation process. For a given output element $p$ within a block, the ideal mantissa-domain dot product is
\begin{equation}
p
=
\sum_{t=1}^{L}
(m_{A})_{t}
(m_{B})_{t},
\end{equation}
where $L$ denotes the size of the dot product determined by the system configuration. We model photonic noise for each mantissa multiplication. Specifically, each product term is perturbed by independent Gaussian noise:
\begin{equation}
\widehat{p}
=
\sum_{t=1}^{L}
\Big(
(m_{A})_{t}
(m_{B})_{t}
+
\epsilon_{t}
\Big),
\qquad
\epsilon_{t}
\sim
\mathcal{N}(\mu, \sigma^2).
\end{equation}

Under this model, the mantissa-domain multiplication error for each output element is given by
\begin{equation}
\Delta p^{\mathrm{pho}}
=
\widehat{p}
-
p
=
\sum_{t=1}^{L}
\epsilon_{t}.
\end{equation}
Since the noise terms are independent, the resulting error also follows a Gaussian distribution:
\begin{equation}\label{eq:gaussian}
\Delta p^{\mathrm{pho}}
\sim
\mathcal{N}\!\left(L\mu,\, L \sigma^2 \right).
\end{equation}
This result shows that photonic multiplication error remains unbiased, while its variance grows linearly with the dot-product length.

After photonic multiplication, partial results are scaled by the combined exponent $E_{A} + E_{B}$ and accumulated in the digital domain.
While exponent recovery does not introduce additional numerical error, it scales the magnitude of existing photonic noise. Consequently, the system error of photonic multiplication of the $k$-th tile after exponent recovery is as follows:
\begin{equation}
\Delta^{sys}
=
2^{E_{A}+E_{B}}
\Delta p^{\mathrm{\textit{pho}}}
\sim
\mathcal{N}\!\left(
2^{(E_{A}+E_{B})}L\mu,\,
2^{2(E_{A}+E_{B})}
L
\sigma^2
\right).
\end{equation}

In LightMat-HP, the bit-width of $E_{A}$ and $E_{B}$ is set by the system configuration. Therefore, we can assume
\begin{equation}
E_{A}+E_{B} \le E_{\max},
\end{equation}
which implies a bound on the combined exponent.
Under the Gaussian model in Equation(\ref{eq:gaussian}), the variance of the system error after exponent recovery for a single output value would be:
\begin{equation}
\mathrm{Var}\!\left(\Delta^{\mathrm{\textit{pho}}}\right)
= 2^{2(E_{A}+E_{B})}L\sigma^2 
\le 
2^{2E_{\max}}L\sigma^2.
\end{equation}
Equivalently, we provide the overall RMS bound on the photonic multiplication error:
\begin{equation}\label{eq:photonic_error}
\sqrt{\mathbb{E}\!\left[{\Delta^{\textit{pho}}}^2\right]}
=
\sqrt{\mathrm{Var}\!\left(\Delta^{pho}\right)}
\le
2^{E_{\max}}\sqrt{L}\;\sigma.
\end{equation}


The above error analysis demonstrates that the numerical accuracy of LightMat-HP is well-bounded and controllable. These properties enable high-precision GEMM and provide a theoretical foundation for the scalability of the LightMat-HP system.

\section{Evaluation Methodology}\label{sec:implementation}
In this section, we describe the evaluation methodology of the proposed LightMat-HP system. 
Section~\ref{subsec:imple_prototype} describes the implementation details of LightMat-HP on our  photonic computing prototype. To evaluate the performance, scalability, and architectural trade-offs of LightMat-HP at system scale, we conduct large-scale simulations that enable comparison with existing hybrid electronic-photonic accelerators as well as state-of-the-art full electronic accelerators. Section~\ref{subsec:large_simulation} details the large-scale simulation framework. Section~\ref{sec:setup} gives the evaluation setup including the workload, evaluation metrics, and baselines for comparison.

\subsection{Implementation of LightMat-HP on Our Photonic Computing Prototype}\label{subsec:imple_prototype}
The hardware architecture of the photonic multiplication prototype is implemented in Vivado 2024.1 using a modular block design approach. The design integrates the processing system (PS) and programmable logic (PL) of the ZCU111 RFSoC through AXI-based interfaces. The ZCU111 evaluation board serves as the reconfigurable platform for the digital subsystem of LightMat-HP, on which key functionalities, including matrix tiling, data serializer, \textit{DAC Players}, \textit{ADC Capture}, and matrix reconstruction are implemented. Each \textit{DAC Player} module includes a BRAM buffer to stream operands to the DACs at high speed. The PYNQ framework is adopted as the software control environment, providing a user interface for matrix input and for obtaining the final computation results, as illustrated in Fig.~\ref{fig:sys_architecture}.

Within the programmable logic, the \textit{DAC Player} and \textit{ADC Capture} modules are constructed using AXI BRAM Controller IPs, Block Memory Generator IPs, and custom RTL logic. The DAC Player reads serialized data from on-chip BRAM and streams it to the RF Data Converter (RFDC), while the ADC Capture module receives digitized data from the RFDC and stores it back into BRAM. We set the size of each BRAM to 32KB, which corresponds to the number of multiplication operations that can be executed per trigger. Synchronization signals are generated and distributed through an AXI GPIO IP to simultaneously trigger the \textit{DAC Players} and \textit{ADC Capture} modules. The AXI control interfaces operate at 100~MHz, while the DAC data streaming clock and ADC capture clock are set to 256~MHz and 512~MHz, respectively. Control communication is handled via AXI interfaces, and the high-throughput data transfer relies on AXI4-Stream interfaces. Both DACs and ADCs operate at a sampling rate of 4.096~GS/s. In addition, calibration of the two MZMs is performed using polynomial fitting on the transfer functions at different orders to achieve accurate photonic multiplication.

\subsection{Large-Scale Simulations}\label{subsec:large_simulation}

We evaluate the proposed LightMat-HP architecture using a large-scale simulation framework that integrates performance, area, and power modeling. This framework combines the off-the-shelf Lightning's large-scale simulations\footnote{The \href{https://github.com/open-photonics/lightning-lts/tree/main/simulation}{\textcolor{blue}{link}} of Lightning large-scale simulations, which reproduces the results reported in Section 9 of the Lightning paper.} and RTL simulations. We modified the Lightning's large-scale simulator to support the end-to-end GEMM process of our architecture, including DRAM and SRAM access latency and bandwidth constraints, scheduler-driven tile dispatching, PPU launch overheads, and parallel execution across multiple PPUs. In addition to performance modeling, the framework incorporates area and power estimation by integrating synthesis-based results for digital logic, memory macro models for on-chip SRAM, system-level models for off-chip DRAM, and reported device-level models for photonic components. By combining cycle-level performance simulation with architecture-level area and power modeling, the framework enables a comprehensive evaluation of LightMat-HP across latency, throughput, energy efficiency, as well as chip area and power.

\subsubsection{Performance Models:} Similar to the photonic computing core in Lightning, we use 97 GS/s DACs and ADCs with 100 GHz modulators and photodetectors in our simulations.  The system clock is set to 1 GHz. LightMat-HP integrates 100 PPUs, each executing GEMM operations for a pair of tiles. And each PPU equips 4 MZMs with the same internal structure (see Figure.\ref{fig:sys_architecture} (b)). The evaluation matrix will be decomposed into tiles, and each PPU will produce a 2 $\times$ 2 output tile using the tile-based matrix multiplication process described in Section~\ref{sec:tile_matrix_multiplication}. Off-chip memory accesses are modeled using a contention-free DRAM abstraction with an effective aggregate bandwidth of 1.5~TB/s, where memory access latency is determined solely by data transfer size and bandwidth. On-chip data movement is modeled using a high-bandwidth SRAM hierarchy, including a shared global SRAM with an effective bandwidth of 6.0~TB/s and per-PPU local SRAMs with an effective bandwidth of 8.0~TB/s, reflecting a highly parallel, banked on-chip memory organization.


The modified simulator explicitly models the computation process of each PPU, including the digital FP-to-BFP/BFP-to-FP conversion and photonic multiplication. The latency of our FP2BFP converter and BFP2FP converter IP and photonic result accumulation are obtained from dedicated RTL simulations, and the photonic computation latency is derived from the photonic domain frequency, the number of elements in each tile to process, and the pulse size of each element.

\subsubsection{Area and Power models:} To evaluate the chip area and power of LightMat-HP, we adopt a joint modeling approach that combines electronic and photonic components. For the electronic part, the digital logic is implemented in RTL and synthesized using Cadence Genus with a 65\,nm standard-cell library \cite{cadence_genus}, including FP/BFP conversion units, dot-product processing units (DPPUs), the scheduler, and the controller. The on-chip SRAM has a total capacity of 41\,MB, consisting of a 16\,MB global shared SRAM and 25\,MB of PPU-local SRAM (100 PPUs, 256\,KB per PPU), which are used to buffer input tiles and store intermediate accumulation results. The area and power of SRAM are estimated using standard memory macro models and synthesis-based estimation. The off-chip DRAM area and energy are modeled according to the system configuration. All electronic component parameters are then scaled to a 7\,nm technology node using scaling equations reported in prior work \cite{stillmaker2017scaling}. For the photonic part, the area and power of the MZM array, photonic detectors, and laser sources are estimated based on reported device models and parameters \cite{peserico2023integrated,tait2022quantifying,zhang2025integrated}. Finally, the electronic and photonic results are combined to obtain the overall area and power breakdown of LightMat-HP, as summarized in Table~\ref{tab:power_area}.

\subsection{Evaluation Setup}\label{sec:setup}

\subsubsection{Matrix Workload}
To assess different application scales and the effectiveness of matrix tiling, we test a set of representative matrix sizes as listed in the Table~\ref{tab:benchmarks}. Each value in the matrices is represented by FP32 format. 


\begin{table}[htbp]
  \centering
  \caption{Matrix sizes and their application scenarios}
  \label{tab:benchmarks}
  \begin{tabular}{|p{0.12\textwidth}|p{0.28\textwidth}|p{0.42\textwidth}|}
    \hline
    \textbf{Category} & \textbf{Sizes} & \textbf{Purpose} \\
    \hline
    \hline
    Small & 16×16, 64×64 & Correctness verification \\ \hline
    Medium & 128×128, 256×256 & AI workloads \\ \hline
    Large & 512×512, 1024×1024 & HPC performance and scalability \\ \hline
  \end{tabular}
\end{table}
\subsubsection{Comparison Baselines}
For comparison, we benchmark LightMat-HP against the following platforms:
\begin{itemize}
    \item \textbf{CPU}: Intel i5-12400F featuring 6 Performance-cores, 12 threads, 2.5GHz clock, and 18MB cache. GEMM is implemented using Intel Math Kernel Library (MKL).
    \item \textbf{GPU}: NVIDIA RTX 3060 featuring 3584 CUDA Cores, 112 Tensor Cores, and 12 GB GDDR6 memory. High-performance matrix multiplication is implemented based on NVIDIA cuBLAS.
    \item \textbf{FPGA}: A custom GEMM kernel synthesized using HLS on the Xilinx ZCU111 RFSoC platform, with computation executed in the programmable logic and controlled by the processing system.
    \item \textbf{BITLUME \cite{chengpeng2025iccad}}: A photonic accelerator with precision-flexible lossless scheme and an optimized round-truncation algorithm that supports multiplications beyond 8-bit precision.
\end{itemize}

\subsubsection{Evaluation Metrics}\label{metrics}
We evaluate LightMat-HP using the following performance metrics. 
\begin{itemize}
\item \textbf{Computational Throughput:} the number of multiplication and addition operations executed per second, measured in giga-operations per second (GFLOPS).
\item \textbf{Latency:} the amount of time for end-to-end execution of a complete matrix multiplication. 
\item \textbf{Energy Efficiency:} the number of multiplication and addition operations executed per watt, measured in GFLOPS/W.
\item \textbf{Relative $\ell_2$ Error (RelL2):} a metric that quantifies the relative error energy of the computed matrix with respect to the electronic reference result \cite{trefethen2022numerical}, defined as:
\begin{equation}
\mathrm{RelL2} = \frac{\left\lVert \mathbf{C}_{\mathrm{measured}} - \mathbf{C}_{\mathrm{ref}} \right\rVert_2}
     {\left\lVert \mathbf{C}_{\mathrm{ref}} \right\rVert_2},
\label{eq:RelL2}
\end{equation}
where $\mathbf{C}_{\mathrm{measured}}$ denotes the output matrix produced by LightMat-HP and $\mathbf{C}_{\mathrm{ref}}$ is the high-precision electronic baseline. 
\item \textbf{Root Mean Square Error (RMSE):} an absolute error metric that captures the overall magnitude of numerical deviations, defined as
\begin{equation}
\mathrm{RMSE} = \sqrt{ \frac{1}{N} \sum_{i=1}^{N} \left( y^{(i)}_{\mathrm{measured}} - y^{(i)}_{\mathrm{ref}} \right)^2 },
\label{eq:RMSE}
\end{equation}
where $N$ is the number of elements in the output matrix. RMSE reflects the accumulation of element-wise errors and provides insight into the absolute noise level introduced by the photonic computing pipeline.
\item \textbf{Mean Absolute Error (MAE):} the average magnitude of element-wise numerical deviations, defined as
\begin{equation}
\mathrm{MAE} =\frac{1}{N} \sum_{i=1}^{N} \left| y^{(i)}_{\mathrm{measured}} - y^{(i)}_{\mathrm{ref}} \right|,
\label{eq:MAE}
\end{equation}
where $N$ is the number of elements in the output matrix. Compared with RMSE, MAE is less sensitive to outliers and provides a more interpretable measure of the typical absolute error per output element.
\end{itemize}

\section{Evaluation Results}\label{sec:evaluation}
We design detailed experiments to evaluate the proposed LightMat-HP system in terms of latency, computational throughput, and energy efficiency. These experiments aim to answer two questions: 
\begin{description}
    \item [Q1:]  How accurately does LightMat-HP compute high-precision matrix multiplication across different dimensions? \item [Q2:] How does LightMat-HP compare with state-of-the-art electronic and photonic accelerators in terms of latency, throughput, and energy efficiency?
\end{description}

\subsection{Experimental Evaluation of Computational Accuracy}
\begin{figure}[thbp]
  \includegraphics[width=0.8\textwidth]{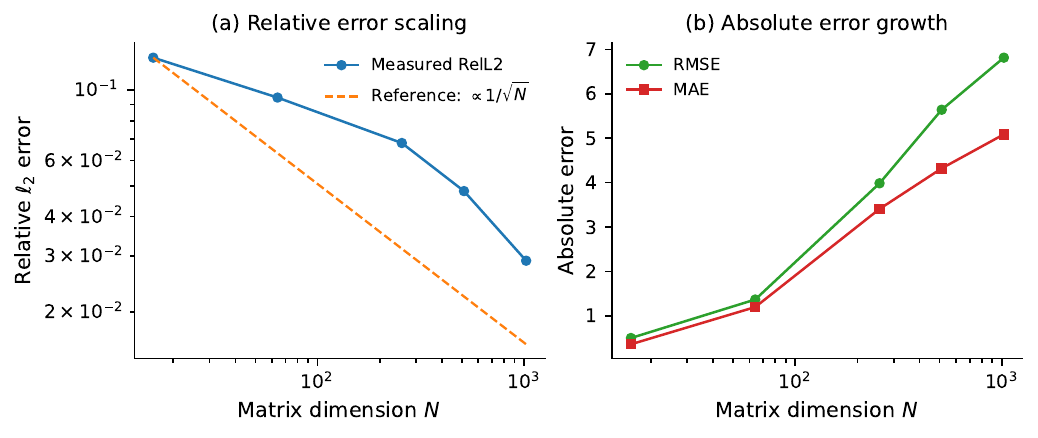}
  \caption{Error Scaling under different matrix dimensions.}\label{fig:exp1_accuracy}
  \Description{}
\end{figure}

Fig.~\ref{fig:exp1_accuracy} shows how the numerical error of LightMat-HP scales with matrix size, including both relative and absolute error metrics. The evaluation covers matrix dimensions ranging from $16\times16$ to $1024\times1024$, and all experiments employ the same tile-based scheduling strategy and photonic–electronic computation pipeline. The results are obtained using a Block Floating Point (BFP) configuration with a shared exponent of 6 bits and a mantissa width of 10 bits, and each mantissa is decomposed into two 5-bit slices for photonic multiplication. 

It can be seen that \textbf{the relative error $\ell_2$ decreases monotonically as the dimension of matrix $N$ increases} and approximately follows a scaling trend $1/\sqrt{N}$. This behavior indicates that the random errors introduced by photonic computation are effectively averaged out in larger matrix multiplications, leading to improved accuracy. In contrast, for small matrices such as $16\times16$, the limited number of accumulated elements provides less statistical averaging, making the relative error more sensitive to element-level deviations, therefore resulting in a higher relative error.

In contrast to the decreasing trend of relative error, the absolute error metrics (RMSE and MAE) gradually increase as the matrix dimension grows. This is because that the number of accumulated terms in matrix multiplication increases with $N$, leading to the accumulation of random noise in absolute magnitude. Nevertheless, the growth of absolute error remains well controlled and does not exhibit superlinear or divergent behavior, indicating that the characteristics of the system noise are stable and predictable.

\begin{figure}[t]
  \centering

  \begin{subfigure}{0.33\textwidth}
    \centering
    \includegraphics[width=\linewidth]{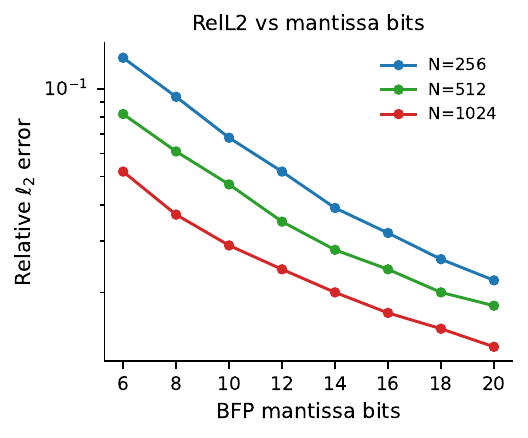}
    \caption{Relative $\ell_2$ error}
    \label{fig:mantissa_relL2}
  \end{subfigure}
  \hfill
  \begin{subfigure}{0.33\textwidth}
    \centering
    \includegraphics[width=\linewidth]{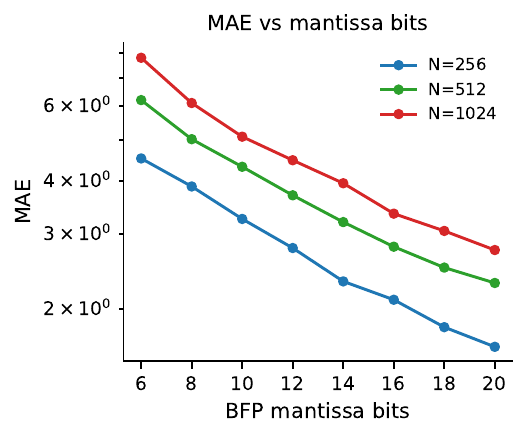}
    \caption{MAE}
    \label{fig:mantissa_mae}
  \end{subfigure}
  \hfill
  \begin{subfigure}{0.33\textwidth}
    \centering
    \includegraphics[width=\linewidth]{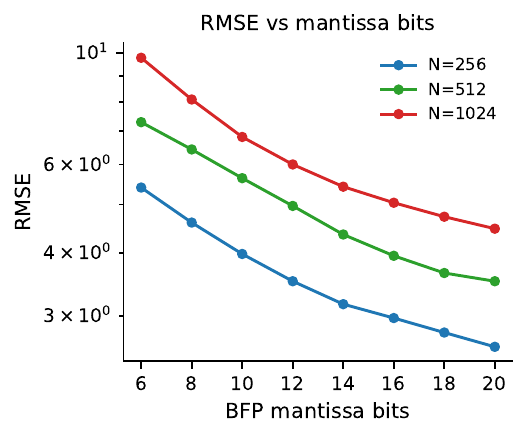}
    \caption{RMSE}
    \label{fig:mantissa_rmse}
  \end{subfigure}

  \caption{Impact of BFP mantissa precision on GEMM accuracy under different matrix dimensions.}
  \label{fig:mantissa_error}
\end{figure}

As shown in Fig.~\ref{fig:mantissa_error}, increasing the BFP mantissa bit width consistently improves GEMM accuracy across all evaluated metrics, including relative $\ell_2$ error, MAE, and RMSE. In this experiment, the mantissa width is swept from 6 to 20 bits. The mantissa is represented in a slice-based form and mapped onto photonic multiplication units at the slice granularity, while the shared exponent configuration remains 6 bits. For a fixed matrix dimension, all error metrics decrease monotonically as the bit width of the mantissa increases, with the most significant improvement observed in the low-precision scope. As mantissa precision continues to increase, the rate of error reduction gradually diminishes, indicating a saturation behavior where errors are no longer dominated by mantissa quantization.

These results demonstrate that the computational accuracy of LightMat-HP exhibits three key properties. First, random errors are effectively averaged as the dimension of the matrix increases, preserving overall computational accuracy. Second, the growth of absolute error remains stable and does not become unbounded with scale. Third, in the same mantissa configuration, larger matrix dimensions result in higher error levels due to error accumulation in large matrix size, while the overall trends remain consistent across different matrix sizes. Together, these observations show that the combination of tile-based matrix multiplication using BFP arithmetic and the intrinsic error characteristics of the photonic computing core enables LightMat-HP to maintain good numerical stability and scalability for large-scale matrix operations, providing a solid foundation for future applications in large-scale AI and scientific computing.

\subsection{Simulation-based Performance Comparison with Baselines}
To provide a comprehensive evaluation of LightMat-HP's performance, we compare it with the baselines mentioned above for matrix multiplication. The evaluation metrics for this experiment include computational throughput (GFLOPS), latency (ms), and energy efficiency (GFLOPS/W).

\begin{figure}[htbp]
  \includegraphics[width=0.6\textwidth]{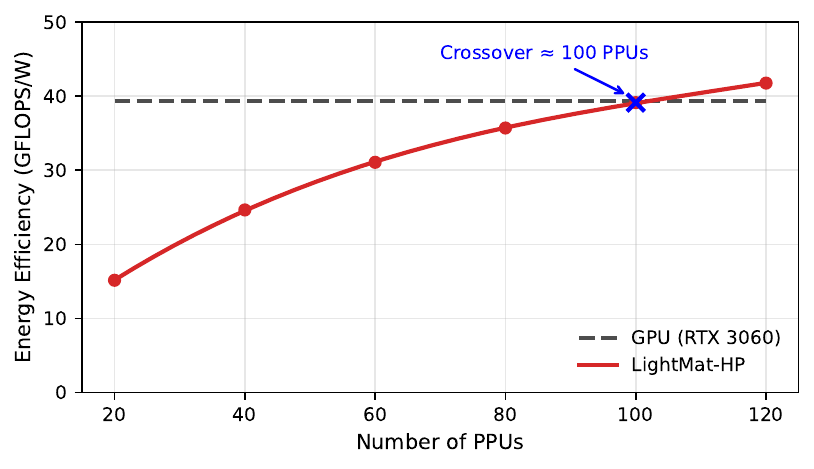}
  \caption{[Simulated] Energy Efficiency Scaling Comparison of LightMat-HP under 1024 $\times$ 1024 matrix multiplication.}\label{fig:num_ppu}
  \Description{}
\end{figure}%
\subsubsection{Scalability} We evaluate how LightMat-HP's energy efficiency evolves as the number of PPU increases in comparison with the NVIDIA RTX~3060 GPU. In this set of simulations, the matrix size is set to $1024\times1024$, and the Block Floating Point (BFP) configuration includes a shared exponent of 6 bits and a mantissa width of 10 bits. As the number of PPUs increases, the system power consumption increases accordingly, and the bandwidth of DRAM and SRAM is scaled to accommodate the increased parallelism as well. As shown in Fig.~\ref{fig:num_ppu}, the energy efficiency of LightMat-HP improves with the increasing number of photonic cores, indicating that photonic matrix multiplication exhibits good inherent parallel scalability. When the number of photonic cores exceeds 100, the simulated result of LightMat-HP achieves higher energy efficiency than NVIDIA RTX~3060 GPU.

\subsubsection{Throughput and Latency}

\begin{figure*}[htbp]
  \includegraphics[width=\textwidth]{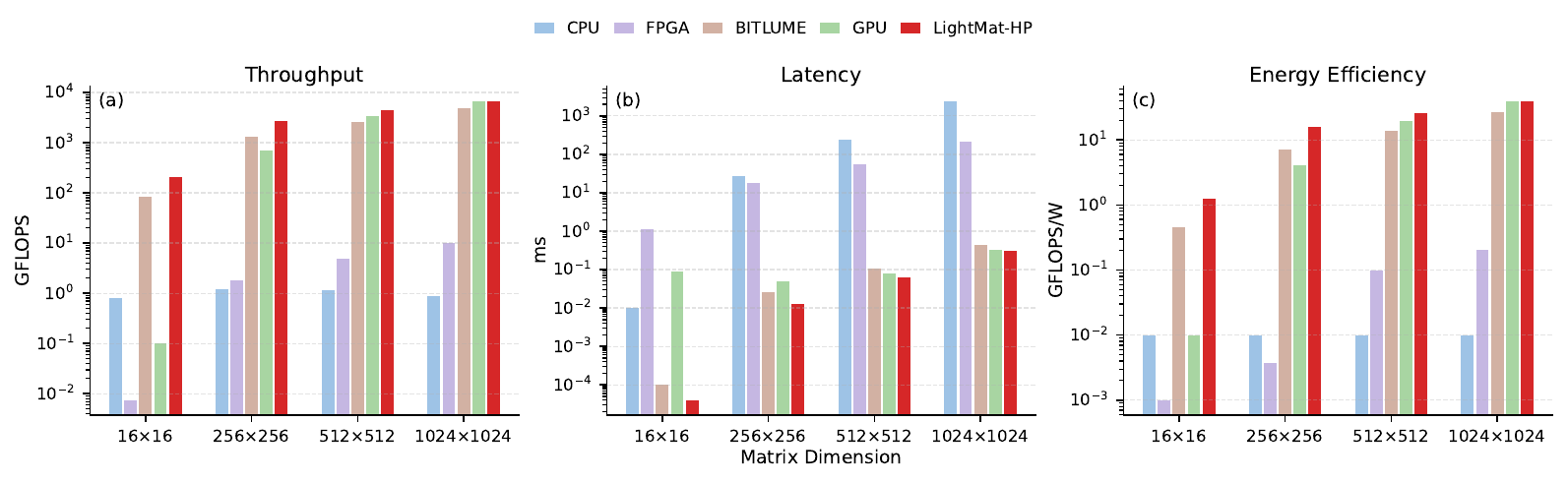}
  \caption{Comparison of latency, throughput, and energy efficiency across different platforms.}\label{fig:exp3_three_metrics}
  \Description{}
\end{figure*}

In these simulations, LightMat-HP is configured with 100 PPUs to achieve energy efficiency comparable to that of the NVIDIA RTX 3060 GPU for matrix size of $1024 \times 1024$. All results are obtained using a BFP configuration with a shared exponent of 6 bits and a mantissa width of 10 bits. Fig.\ref{fig:exp3_three_metrics} compares the throughput and the latency of LightMat-HP with electronic computing platforms (CPU, GPU, FPGA) as well as the state-of-the-art photonic accelerator (BITLUME\cite{chengpeng2025iccad}). Across all matrix sizes from small size ($16 \times 16$) to large size ($1024 \times 1024$), LightMat-HP consistently achieves higher throughput and lower latency than the CPU, FPGA, BITLUME, and GPU, especially at small and medium matrix sizes. As the matrix size increases, the throughput of LightMat-HP scales from 209.25~GFLOPS to a peak of 6654.44~GFLOPS, demonstrating the effectiveness of our architecture and the photonic parallelism.

In terms of latency, LightMat-HP outperforms CPU, GPU and FPGA across all evaluated matrix sizes. Compared with BITLUME, the corresponding speedups are $2.48\times$, $2.05\times$, $1.70\times$, and $1.36\times$. These results indicate that LightMat-HP delivers lower latency than BITLUME for all the small ($16\times16$) and medium-sized ($256\times256$) matrix multiplications, while providing larger performance gains over CPU and FPGA baselines. Similar trends persist at larger scales ($512\times512$ and $1024\times1024$). This advantage comes from the overall architectural design of LightMat-HP, including the optical signal multiplexing and the BFP arithmetic for matrix multiplication.

The observed latency differences are the result of different execution models, the CPU baseline is limited by sequential execution and memory access. As matrix size increases, throughput stays nearly constant while latency grows rapidly, making CPUs poorly suited for large matrix multiplication. GPU (RTX3060) achieves higher throughput for large matrices because of massive parallelism, but small matrices suffer from kernel launch and scheduling overhead. As a result, GPUs are better suited for large-scale workloads but are less effective for small or latency-sensitive tasks. For FPGAs, increasing matrix size improves pipeline utilization and leads to higher throughput; nevertheless, fixed pipeline depth, data transfer overhead, and PS--PL coordination costs limit further improvements in both throughput and latency compared to photonic architectures. BITLUME exhibits only modest throughput gains as matrix size increases, while its latency rises sharply, indicating that scheduling overhead limits its overall acceleration effectiveness.

In contrast, both BITLUME and LightMat-HP exploit the inherent parallelism of photonic computation, enabling low computation latency at small matrix sizes. However, BITLUME relies on more rigid dataflow and accumulation mechanisms, leading to increasing overheads from data movement, optical-to-electrical conversion, and partial-sum accumulation as matrix sizes grow. LightMat-HP alleviates these bottlenecks through its system-level architectural design, including optical signal multiplexing and efficient slice-based BFP arithmetic. As a result, LightMat-HP sustains lower latency across all scales, with particularly pronounced advantages at small and medium sizes, while maintaining consistent gains at larger matrix sizes.

For energy efficiency, it can be seen that, as the matrix size increases, the energy efficiency of LightMat-HP improves from 1.23 GFLOPS/W for the $16\times16$ matrix to 39.14 GFLOPS/W for the $1024 \times 1024$ matrix, indicating that the fixed overhead in the system (such as optoelectronic interfaces, control logic, and laser) can be amortized over large-scale computations. For the $1024 \times 1024$ matrix multiplication workload, LightMat-HP outperforms CPU, FPGA, and BITLUME in energy efficiency, and also maintains a competitive advantage in throughput and end-to-end latency. Compared to the SOTA photonic accelerator BITLUME, LightMat-HP achieves a $1.48\times$ improvement in energy efficiency. Furthermore, LightMat-HP is close to the GPU (RTX3060) in energy efficiency (39.41 GFLOPS/W), its energy efficiency can be improved by further increasing the number of PPUs and bandwidth, combining the optimization methods. These results demonstrate that LightMat-HP exhibits good potential in energy efficiency.

\begin{table}[t]
\centering
\caption{Electronic and Photonic Device Parameters of LightMat-HP}
\label{tab:power_area}
\renewcommand{\arraystretch}{1.15}
\setlength{\tabcolsep}{6pt}
\begin{tabular}{c l c c c c}
\toprule
 & \textbf{Devices} 
 & \textbf{Area (mm$^2$)} 
 & \textbf{Portion} 
 & \textbf{Power (W)} 
 & \textbf{Portion} \\
\midrule

\multirow{7}{*}{\textbf{Electronic}}
 & DRAM & 74.37 & 6.28\%   & 7.42  & 4.35\% \\
 & SRAM & 15.96 & 1.35\%  & 8.02  & 4.70\% \\
 & Digital Logic (FP/BFP, DPPU) & 63.74 & 5.38\% & 36.80 & 21.60\% \\
 & Scheduler  & 9.74  & 0.82\% & 4.91  & 2.88\% \\
 & Controller & 8.49  & 0.72\% & 5.68  & 3.33\% \\
 & DACs & 430.66 & 36.39\% & 52.85 & 31.00\% \\
 & ADCs & 231.73 & 19.57\% & 33.17 & 19.46\% \\
\midrule
\multirow{3}{*}{\textbf{Photonic}}
 & MZM array                    
 & 344.94 & 29.48\% & --     & --      \\
 & Photonic detector & 0.20   & 0.02\%  & 1.53  & 0.90\%  \\
 & Laser & 0.03   & 0.00\% & 20.03 & 11.75\% \\
\midrule
\textbf{Total} & 
 & \textbf{1183.86} & \textbf{100.00\%} 
 & \textbf{170.47} & \textbf{100.00\%} \\
\bottomrule
\end{tabular}
\end{table}

\subsubsection{Area and Power} 

Table~\ref{tab:power_area} summarizes the area and power breakdown of the electronic and photonic components of the LightMat-HP system. It should be noted that the reported area and power parameters are derived from architecture-level modeling and normalized estimation, with the goal of characterizing the relative resource distribution across major system components. From the perspective of the electronic subsystem, LightMat-HP is composed of memory components, digital logic, and optoelectronic interface modules. DRAM and SRAM occupy 74.37~mm$^2$ (6.28\%) and 15.96~mm$^2$ (1.35\%) of the total chip area, respectively, with power consumptions of 7.42~W (4.35\%) and 8.02~W (4.70\%). Although SRAM occupies a relatively small fraction of the total area, its power consumption is comparable to that of DRAM due to its high-bandwidth, multi-ported access characteristics. The digital logic components, including the FP/BFP conversion modules and the DPPU unit, occupy 63.74~mm$^2$ (5.38\%) of the total area while consuming 36.80~W, accounting for 21.60\% of the overall power budget. This observation indicates that, although the core matrix multiplication is performed in the photonic domain, a substantial amount of digital logic is still required to support precision management, system control, and result reconstruction. In addition, the scheduler and controller occupy 9.74~mm$^2$ and 8.49~mm$^2$ of area, respectively, with power consumptions of 4.91~W and 5.68~W. Although their contributions to the total area and power are relatively modest, these components play a critical role in coordinating multi-PPU execution and dataflow scheduling. Unlike existing photonic accelerators, the optoelectronic interface dominates the area and power consumption of LightMat-HP. In particular, the DAC modules occupy 430.66~mm$^2$, accounting for 36.39\% of the total area, and consume 52.85~W (31.00\% of the total power), while the ADC modules occupy 231.73~mm$^2$ (19.57\%) and consume 33.17~W (19.46\%). These results indicate that analog-to-digital (AD) and digital-to-analog (DA) conversion remain major bottlenecks in hybrid photonic--electronic computing architectures. This behavior is not unique to LightMat-HP but reflects a common limitation of high-performance photonic computing platforms. In the photonic subsystem, the MZM array occupies an area of 344.94~mm$^2$, accounting for 29.48\% of the total system area, making it the largest photonic component in terms of physical footprint. This is primarily due to the size constraints of currently manufacturable photonic devices. In this work, the MZM array is treated as a passive component, whose power consumption is dominated by the associated front-end DACs and the laser source; therefore, its power is not explicitly listed in the device parameters. The photonic detector occupies only 0.20~mm$^2$ and consumes 1.53~W, having a negligible impact on the overall system area and power. In contrast, the laser source occupies a minimal area of 0.03~mm$^2$ but consumes 20.03~W, accounting for 11.75\% of the total system power, and thus significantly influences the overall energy consumption of the system.

Overall, the LightMat-HP system has a total area of 1183.86~mm$^2$ and a total power consumption of 170.47~W. Its area and power consumption distribution are consistent with the characteristics of a hybrid optoelectronic computing system: \textbf{the photonic computing core has high energy efficiency in matrix multiplication, while the system overhead is mainly determined by the optoelectronic interface modules and digital logic.} As photonic device sizes further shrink and the efficiency of AD/DA conversion continues to improve, this system overhead is expected to gradually decrease, thereby further enhancing the throughput and energy efficiency of the photonic accelerator.

\section{Related Work}\label{sec:relatedwork}
\subsection{Photonic Matrix Multiplication}
Most photonic approaches for matrix multiplication focus on multiply-accumulate (MAC) and matrix-vector multiplication (MVM) operations, which naturally map to the linear superposition of optical signals and can be composed into GEMM at the system level~\cite{cheng2021photonic,peserico2023integrated}. One class of these approaches uses MZI meshes to implement linear computation, often based on matrix factorization techniques such as singular value decomposition~\cite{clements2016optimal,shen2017deep}. These architectures offer high programmability and theoretical generality; however, they require precise phase control and frequent calibration, which introduces significant control overhead and limits scalability~\cite{peserico2023integrated}. Another class of methods relies on WDM, where matrix weights are encoded using wavelength-selective components such as microring resonators or weight banks, enabling highly parallel broadcast-and-weight operations with high energy efficiency~\cite{xu202111}. However, thermal sensitivity, wavelength drift, and inter-channel crosstalk make weight stabilization and numerical precision challenging at scale. The third class of methods exploits spatial-domain optical processing, such as multi-plane light conversion or diffractive optics, to realize large-scale linear transformations with high spatial parallelism~\cite{cheng2021photonic}. However, these systems typically offer limited reconfigurability and face challenges in tight electronic-photonic integration and dynamic weight updates.

\subsection{High-Precision Photonic Computing}
High numerical precision remains one of the most challenging issues in photonic computing systems. Unlike electronic digital circuits, photonic computing relies on analog physical processes such as optical interference, modulation, and photo-detection, which are highly sensitive to device noise and environmental variations. As a result, early research mainly focused on demonstrating the feasibility of performing matrix operations rather than achieving high numerical precision. Such early photonic systems typically provided only 4--8 bits of effective precision and relied heavily on device calibration, making them more suitable for inference workloads that do not require high numerical accuracy~\cite{shen2017deep, hamerly2019large}. To improve computational accuracy, subsequent studies introduced calibration and error-compensation mechanisms, including device tuning, repeated optical computations with averaging, and electronic post-processing. 
Under controllable experimental conditions, these methods can increase the effective precision to the range of 7--10 bits~\cite{tait2017neuromorphic, feldmann2019all, ahmed2025universal}. For example, \cite{hua2025integrated} enhanced the effective precision to 7.61 bits by accurately recovering and compensating the amplitude and phase responses of a programmable photonic integrated circuit using on-chip reference paths and self-calibration algorithms. With high-resolution ADCs and digital post-processing, some systems can further achieve end-to-end numerical accuracy comparable to FP16 or even FP32~\cite{demirkiran2024mirage, chengpeng2025iccad}.



\subsection{Hybrid Photonic-Electronic Architecture}
Although pure photonic computing has been studied for decades~\cite{xiang2021review}, most practical systems adopt hybrid photonic-electronic architectures~\cite{feldmann2021parallel,rios2019memory}. This reflects that essential functions, including logic control, nonlinear activation, memory access, and synchronization, are still more efficiently handled in the electronic domain. As a result, a common paradigm is to use photonic modules for high-throughput linear operations such as matrix multiplication, while delegating control flow and nonlinear processing to electronic components~\cite{zhong2023lightning,zhang2025rocket}. MIT researchers developed Lightning, a reconfigurable photonic-electronic SmartNIC for deep learning inference, and it decouples the compute control plane from the data plane using the count-action abstraction~\cite{zhong2023lightning}. However, this count-based synchronization aligns operands at the granularity of each individual multiplication, introducing substantial electronic–photonic coordination overhead. ADEPT performs GEMM operations with a photonic computing unit and non-GEMM operations using a vectorized digital electronic ASIC~\cite{demirkiran2023electro} while its performance gains depend on carefully balanced workload partitioning and remain sensitive to electronic control and memory access latency. Mirage combines photonic MAC units with residue number systems to support training workloads~\cite{demirkiran2024mirage}, but the required multi-modulus datapaths and frequent digital-domain conversions incur substantial area and energy overhead. More recent designs, such as ROCKET and BITLUME, reduce the number of photonic components and improve data reuse~\cite{zhang2025rocket,chengpeng2025iccad}, yet still rely on carefully orchestrated electronic pipelines to match photonic throughput. All these studies target accelerating deep learning inference or training, and none of them provides a general, system-level framework for accelerating general matrix multiplication. 

\section{Conclusions and Future Work}\label{sec:conclusion}
This work proposes LightMat-HP, a novel hybrid photonic-electronic computing system designed to support high-precision general matrix multiplication acceleration. The experimental studies on the photonic-electronic testbed motivate our design of tiled-based GEMM for mantissa matrices using BFP arithmetic. Experimental results demonstrate that the LightMat-HP achieves a peak performance with throughput of $> 6600$ GFLOPS and an energy efficiency of 39.14 GFLOPS/W, outperforming the electronic baselines and SOTA photonic accelerators. The statistical trend of relative errors and absolute errors demonstrates stable computational accuracy across different matrix sizes. These results confirm the feasibility and efficiency of LightMat-HP architecture and demonstrate that hybrid photonic and electronic architectures can offer high performance, energy efficiency, and reliable precision, providing a promising foundation for future computing systems that aim to support large-scale artificial intelligence and scientific applications.

For future work, LightMat-HP can be scaled into a high-performance photonic computing cluster composed of multiple interconnected accelerators. This system-level evolution allows LightMat-HP to be applied to classical and emerging workloads, such as large language models training and inference, as well as scientific computing applications, where massive parallelism, high precision, and energy efficiency are simultaneously required.


\bibliographystyle{ACM-Reference-Format}
\bibliography{refs}












\end{document}